\documentclass[prd,twocolumn,preprintnumbers,superscriptaddress,showpacs]{revtex4-1}
\usepackage[colorlinks=true, pdfstartview=FitV, linkcolor=red, citecolor=blue, urlcolor=blue, pdftitle={},pdfsubject={}, pdfkeywords={}]{hyperref}

\usepackage{amsmath,amssymb}
\usepackage{graphicx}
\usepackage{dcolumn}
\usepackage{bm}
\usepackage{slashed}
\usepackage{marvosym}
\usepackage{textcomp}
\usepackage{threeparttable}
\usepackage{booktabs}
\usepackage{setspace}
\usepackage{latexsym}
\usepackage{amsmath}
\usepackage{mathrsfs}
\renewcommand{\TPTtagStyle}%
{\normalsize\textit}
\usepackage{color}

\newcommand{\blue}[1]{{\color{blue} #1} }

\newcommand{\nn}{\nonumber}

\allowdisplaybreaks[1]

\usepackage[normalem]{ulem}

\begin{document}
\title{Axial Kinetic Theory and Spin Transport for Fermions with Arbitrary Mass}
\author{Koichi Hattori}
\affiliation{
	Yukawa Institute for Theoretical Physics, Kyoto University, Kyoto 606-8502, Japan.\\
}
\author{Yoshimasa Hidaka}
\affiliation{
	RIKEN Nishina Center,
	RIKEN, Wako, Saitama 351-0198, Japan.\\
}
\affiliation{
	RIKEN iTHEMS, RIKEN, Wako, Saitama 351-0198, Japan.
}
\author{Di-Lun Yang}
\affiliation{
	Yukawa Institute for Theoretical Physics, Kyoto University, Kyoto 606-8502, Japan.\\
}
\affiliation{Faculty of Science and Technology,  Keio University, Yokohama 223-8522, Japan}
\preprint{RIKEN-QHP-408}
\preprint{YITP-19-14}
\begin{abstract}
We derive the quantum kinetic theory for fermions with arbitrary mass in a background electromagnetic field 
from the Wigner-function approach.
Since spin of massive fermions is a dynamical degree of freedom, 
the kinetic equations with the leading-order quantum corrections describe entangled dynamics of 
not only the vector- and axial-charge distributions but also of the spin polarization. 
Therefore, we obtain one scalar and one axial-vector kinetic equations with 
magnetization currents pertinent to the spin-orbit interaction. 
We show that our results smoothly reduce to the massless limit 
where the spin of massless fermions is no longer an independent dynamical degree of freedom 
but is enslaved by the chirality and momentum and the accordingly kinetic equations turn into the chiral kinetic theory for Weyl fermions. 
We provide a kinetic theory covering both the massive and massless cases, and 
hence resolves the problem in constructing the bridge between them. 
Such generalization may be crucial for applications to various physical systems.
Based on our kinetic equations, we discuss the anomalous currents 
transported by massive fermions in thermal equilibrium.  
\end{abstract}

\maketitle

\section{Introduction}
Triggered by predictions of the chiral magnetic/vortical effect (CME/CVE) \cite{Vilenkin:1979ui, Kharzeev:2007jp,Fukushima:2008xe}, 
the transport of Weyl fermions is widely studied in recent years. 
In light of connections to quantum anomalies, those transport phenomena 
have attracted much attention in systems with quite different energy scales 
including relativistic heavy-ion collisions \cite{Kharzeev:2015znc, Hattori:2016emy}, 
Weyl semimetals \cite{Li:2014bha}, and lepton transport in supernovae explosions \cite{Yamamoto:2015gzz,Masada:2018swb}.

To investigate such anomalous transport in out-of-equilibrium systems, 
the chiral kinetic theory (CKT) has been developed to capture 
the chiral anomaly effects \cite{Gao:2012ix, Son:2012wh, Stephanov:2012ki, Son:2012zy,Chen:2012ca,
Manuel:2013zaa,Chen:2014cla,Chen:2015gta,Hidaka:2016yjf,Hidaka:2017auj,Hidaka:2018ekt,
Mueller:2017arw,Mueller:2017lzw,Huang:2018wdl,Carignano:2018gqt,Dayi:2018xdy,Liu:2018xip,Lin:2019ytz}. 
Particularly, recent progresses constructed a robust bridge between the CKT and quantum field theory 
on the basis of the $ \hbar $ expansion applied to the Wigner functions, 
which allows for systematic derivation of the side-jump effects 
stemming from the spin-orbit coupling and collisions \cite{Chen:2014cla,Hidaka:2016yjf,Hidaka:2017auj}.

However, the CKT developed for massless fermions appears to have an issue in its connection 
to the existing quantum kinetic theory for massive fermions \cite{Heinz:1983nx,Elze:1986hq,Vasak:1987um,Blaizot:2001nr,Mueller:2019gjj}. 
There are crucial differences between the massless and massive fermions 
as representations of the Lorentz symmetry. 
Whereas spin of Weyl fermion is enslaved by its momentum and is not an independent dynamical degree of freedom, 
spin of massive Dirac fermions is subject to dynamical effects. 
It is, thus, necessary to understand how the side jumps and magnetic-moment coupling 
in CKT are reduced from the dynamics of massive fermions to the massless limit.

In the aforementioned systems, mass effects will play sizable roles. 
For example, the measurements of global polarization for $\Lambda$ hyperons in heavy-ion collisions \cite{STAR:2017ckg,Adam:2018ivw} motivated by theoretical predictions \cite{Liang:2004ph,Becattini:2013vja} have triggered increasing studies upon the spin-polarization formation and angular momenta of relativistic fluids \cite{Becattini2013a,Fang:2016vpj,Florkowski:2017ruc,Yang:2018lew,Fukushima:2018osn,Florkowski:2018fap,Florkowski:2018ahw,Hattori:2019lfp}.  
Since the spin of $\Lambda$ is mainly attributed to the strange-quark component, 
one may not treat them as massless fermions as compared to temperature of the quark-gluon plasma. 
In addition, the mass corrections upon the axial currents, generated by axial-CVE 
and chiral separation effect (CSE), has accordingly received further attentions \cite{Gorbar2013,Buzzegoli:2017cqy,Lin:2018aon}. 
As for the astrophysical applications of the chiral-plasma instability \cite{Akamatsu:2013pjd}, 
a critical question was raised in the relaxation time of axial charge due to 
effects of electron mass \cite{Ohnishi:2014uea,Grabowska:2014efa}. 
They remain open questions and will be important applications of the CKT with the mass correction,
which can simultaneously trace the time evolution of the charge transport, chiral imbalance, and spin polarization.

In this paper, we apply the Wigner-function approach to derive a quantum kinetic theory for fermions with arbitrary mass,
which we call the axial kinetic theory (AKT).
Recently, related studies were presented in Refs.~\cite{Weickgenannt:2019dks,Gao:2019znl}, 
in which the kinetic theories are derived in the rest frame of massive fermions.
Although physics is frame invariant (analogous to gauge invariance), the choice of rest frame similar to the choice of a particular gauge is legitimate only for fermions with mass larger than typical electromagnetic and gradient scales. Physically, one simply cannot define a rest frame for massless particles. The kinetic theories derived therein consequently causes divergence and the breakdown of $\hbar$ expansion for smaller mass.
In order to apply a relativistic situation such as heavy-ion collisions, where the quark mass is small or comparable to gradient scales,
one needs the theory applicable to an arbitrary frame (or at least a proper frame). The AKT covers both the massive and massless cases, and 
hence resolves the problem in constructing the bridge between them and should be regarded as the underlying theory which embodies the effective theories obtained in Refs.~\cite{Weickgenannt:2019dks,Gao:2019znl} for the large-mass regime.  
After the formulation, we discuss the anomalous currents transported by massive fermions in a thermal equilibrium, 
which are important in heavy-ion collisions and neutron-star physics. 

This paper is organized in the following order. In Sec.II, we present the master equations obtain from the Wigner-function approach. 
In Se.III, the perturbative solution of the vector part for WFs is derived and a corresponding scalar kinetic equation in AKE is obtained. 
In Sec.IV, we further derive the axial part of WFs partially with an alternative approach and 
generalization, where we also present a corresponding axial-vector kinetic equation in AKE. 
In Sec.V, we 
discuss anomalous transport in thermal equilibrium in our formalism. We then make brief conclusions and outlook in Sec.VI. The details of derivations and computations are presented in Appendices.

\section{Wigner functions and Master Equations}
We consider a massive Dirac field $ \psi $ which is, unlike a massless Dirac field, 
no longer decomposed into a pair of Weyl fermions. 
The Wigner transformation applied to quantum expectation values of correlation functions reads 
\begin{eqnarray}
\grave{S}^{<(>)}(q,X)=\int d^4Ye^{\frac{iq\cdot Y}{\hbar}}S^{<(>)}(x,y),
\end{eqnarray}
where $X=(x+y)/2$ and $Y=x-y$. 
Here, $S^<(x,y)=\langle\bar{\psi}(y)\psi(x)\rangle$ and $S^>(x,y)=\langle\psi(x)\bar{\psi}(y)\rangle$ 
are lessor and greater propagators, respectively. Hereafter, we focus on $S^<(x,y)$. 
Note also that the gauge link is implicitly embedded and $q^{\mu}$ thus represents the kinetic momentum. 
We then apply the decomposition based on the Clifford algebra \cite{Vasak:1987um},
\begin{eqnarray}
\grave{S}^<=\mathcal{S}+ i\mathcal{P}\gamma^5+ \mathcal{V}^{\mu}\gamma_\mu+\mathcal{A}^{\mu}\gamma^5\gamma_{\mu}+ \frac{\mathcal{S}^{\mu\nu}}{2}\Sigma_{\mu\nu},
\end{eqnarray}
where $\Sigma_{\mu\nu}=i[\gamma_{\mu},\gamma_{\nu}]/2$ and $\gamma^5=i\gamma^0\gamma^1\gamma^2\gamma^3$.
The coefficients $\mathcal{V}^{\mu}$ and $\mathcal{A}^{\mu}$ contribute to the vector and axial charge currents, while $\mathcal{S}$ and $\mathcal{P}$ are related to quark and chiral condensates, respectively. 
The antisymmetric $\mathcal{S}^{\mu\nu}$ is related to magnetization. 

For simplicity, we work in the regime where the collision effects are sufficiently weak 
and drop the contribution from the self-energy. Then the lessor propagator obeys
\begin{equation}
\begin{split}\label{eq:masterEq}
(\slashed{\Pi}-m)\grave{S}^{<} 
+\gamma^{\mu}i\frac{\hbar}{2}\nabla_{\mu}\grave{S}^{<}=0,
\end{split}
\end{equation}
where $m$ is the mass of the fermion, 
$ \nabla_{\mu}=\Delta_{\mu}
+\mathcal{O}(\hbar^2)$, 
$\Pi_{\mu}=q_{\mu}+\frac{\hbar^2}{12}(\partial_{\rho}F_{\nu\mu})\partial^{\rho}_q\partial^{\nu}_{q}+\mathcal{O}(\hbar^4)$, 
and $\Delta_{\mu}=\partial_{\mu}+F_{\nu\mu}\partial^{\nu}_{q}$ 
with $F_{\mu\nu}$ being a background-field strength. 
Equation~\eqref{eq:masterEq} can be written into 10 equations with 32 degrees of freedoms~\cite{Vasak:1987um}. 
Three of them read 
\begin{align}
&m\mathcal{S}=\Pi\cdot\mathcal{V},
\quad
m\mathcal{P}=-\frac{\hbar}{2}\nabla_{\mu}\mathcal{A}^{\mu},
\nn
\\
&m\mathcal{S}_{\mu\nu}=-\epsilon_{\mu\nu\rho\sigma} \Pi^{\rho}\mathcal{A}^{\sigma}
+\frac{\hbar}{2}\nabla_{[\mu}\mathcal{V}_{\nu]}
,
\end{align}
where $A_{[\mu}B_{\nu]}\equiv A_{\mu}B_{\nu}-B_{\nu}A_{\mu}$. 
Therefore, one can choose either eight functions $\mathcal{S}$, $\mathcal{P}$, 
and $\mathcal{S}^{\mu\nu}$ as a set of independent functions, 
or the other half $\mathcal{V}^{\mu}$ and $\mathcal{A}^{\mu}$~\cite{Ochs:1998qj}. 
We choose the latter set and apply an $\hbar$ expansion to the rest of equations, 
which results in 
\begin{eqnarray}\label{eqv1}
\Delta\cdot\mathcal{V}&=&0,
\\\label{eqv2}
(q^2-m^2)\mathcal{V}_{\mu}&=&-\hbar\tilde{F}_{\mu\nu}\mathcal{A}^{\nu}_,
\\\label{eqv3}
q_{\nu}\mathcal{V}_{\mu}-q_{\mu}\mathcal{V}_{\nu}
&=&\frac{\hbar}{2}\epsilon_{\mu\nu\rho\sigma}\Delta^{\rho}\mathcal{A}^{\sigma},
\\\label{eqA1}
q\cdot\mathcal{A}&=&0,
\\\label{eqA2}
(q^2-m^2)\mathcal{A}^{\mu}&=&\frac{\hbar}{2}\epsilon^{\mu\nu\rho\sigma}q_{\sigma}
\Delta_{\nu}\mathcal{V}_{\rho}
,
\\\label{eqA3}
q\cdot\Delta\mathcal{A}^{\mu}+F^{\nu\mu}\mathcal{A}_{\nu}
&=&
\frac{\hbar}{2}\epsilon^{\mu\nu\rho\sigma}(\partial_{\sigma}F_{\beta\nu})\partial^{\beta}_{q}\mathcal{V}_{\rho},
\end{eqnarray}     
where $\tilde{F}^{\mu\nu}=\epsilon^{\mu\nu\alpha\beta}F_{\alpha\beta}/2$. 
We have retained the leading-order quantum corrections, 
and removed one redundant equation which can be reproduced from the above set, where the detailed derivations are shown in Appendix~\ref{sec:DerivationOfMasterEquations}.

\section{Vector Wigner functions/scalar kinetic equation}
We now seek perturbative solutions $(\mathcal{V}/\mathcal{A})^{\mu} = (\mathcal{V}/\mathcal{A})^{\mu}_{0}+\hbar (\mathcal{V}/\mathcal{A})^{\mu}_{1}$ up to $\mathcal{O}(\hbar^1)$. 
The zeroth-order solutions are immediately obtained from Eqs.~(\ref{eqv2})-(\ref{eqA2}) as 
\begin{align}\label{LO_WF}
(\mathcal{V}_{0}/\mathcal{A}_0)^{\mu}=2\pi (q/a)^{\mu}\delta(q^2-m^2)f_{V/A},
\end{align}
where $f_{V/A}(q,X)$ represent the vector/axial distribution functions. 
Here, $a^{\mu}(q,X)$ satisfies $q\cdot a=q^2-m^2$
and corresponds to the (non-normalized) spin four vector. 
As shown below, we have $a^{\mu}=q^{\mu}$ in the massless limit 
because the spin is enslaved by the momentum. 
However, $a^{\mu}$ is a dynamical variable in the massive case 
which should be determined by the kinetic theory.

Hence, we anticipate to derive the  scalar kinetic equation (SKE)
and  axial-vector kinetic equation (AKE)  governing 
the dynamical degrees of freedoms $f_{V/A}$ and $a^{\mu}$, 
Plugging Eq.~(\ref{LO_WF}) into Eqs.~(\ref{eqv1}) and (\ref{eqv3}), 
one acquires the LO kinetic equations, $\delta(q^2-m^2)q\cdot\Delta f_V=0$ 
and $\delta(q^2-m^2)\Box_{\mu\nu}\tilde{a}^{\nu}=0$, 
where $\tilde{a}^{\mu}=a^{\mu}f_A$ and $\Box_{\mu\nu}\tilde{a}^{\nu}=q\cdot\Delta\tilde{a}_{\mu}+F_{\nu\mu}\tilde{a}^{\nu}$. The spin part is the renown Bargmann-Michel-Telegdi equation \cite{BLT_spin}.

For $\mathcal{O}(\hbar^1)$ solutions, we first focus on the vector part, 
which can be derived from Eqs.~(\ref{eqv1})-(\ref{eqv3}).  
Similar to the massless case \cite{Hidaka:2016yjf,Huang:2018wdl}, 
Eqs.~(\ref{eqv2}) and (\ref{eqv3}) determine the modification of the dispersion relation and the magnetization-current (MC) term, respectively. 
Accordingly, we find \footnote{Similar to the massless case \cite{Hidaka:2016yjf}, the $\mathcal{O}(\hbar^1)$ terms proportional to $q^{\mu}\delta(q^2-m^2)$ can be absorbed into the distribution functions.}
\begin{eqnarray}
\label{Y_sol_V1}
\mathcal{V}_{1}^{\mu}&=&2\pi\tilde{F}^{\mu\nu}a_{\nu}\delta'(q^2-m^2)f_A+ 2\pi\delta(q^{2}-m^{2}) G^{\mu},
\\
\label{SJ-V}
G^{\mu}
&=&
\frac{ \epsilon^{\mu\nu\rho\sigma}n_{\nu}}{2q\cdot n}
[ \Delta_{\rho}(a_{\sigma}f_A)+F_{\rho\sigma}f_A]
,
\end{eqnarray} 
where $\delta'(q^{2}-m^{2})\equiv d\delta(q^{2}-m^{2})/dq^{2}$, and
$n^{\mu}(X)$ corresponds to a local frame vector specifying the spin basis. See Appendix~\ref{eq:PerturbativeSolutionForWignerFunctions} for more details of the derivation. The presence of MC term implies that $f_V$ is frame dependent, which follows the modified frame transformation between arbitrary frames $n^{\mu}$ and $n'^{\mu}$, 
\begin{eqnarray}
f^{(n')}_V-f_V^{(n)}=\frac{\hbar\epsilon^{\lambda\nu\rho\sigma}n_{\lambda}n'_{\nu}}{2(q\cdot n)(q\cdot n')}\big(\Delta_{\rho}a_{\sigma}+F_{\rho\sigma}\big)f_A,
\end{eqnarray} 
as derived in Appendix~\ref{sec:FrameIndependence}, where the superscripts $(n')/(n)$ of $f_V$ denote the frame dependence. Note that $f_{V/A}$ are frame independent at $\mathcal{O}(\hbar^0)$. 
When one defines the spin basis in the massive particle's rest frame, 
the explicit form of the frame vector reads $n^{\mu}=q^{\mu}/m$ such that $ q \cdot n = m $, 
and the above expressions reduce to those obtained in Ref.~\cite{Weickgenannt:2019dks}, whereas this frame choice is only valid at large mass when $m\grave{S}^<\gg |\gamma\cdot\Delta \grave{S}^<|$. It is necessary to choose a different frame for smaller mass. See also Appendix~\ref{sec:Spin-Hall Effect} for further discussions upon this issue.

When $m=0$ and $a^{\mu}=q^{\mu}$, $G^{\mu}$ reproduces the side-jump term for massless fermions \cite{Chen:2014cla,Hidaka:2016yjf}.
Inserting Eqs.~(\ref{LO_WF})-(\ref{SJ-V}) into Eq.~(\ref{eqv1}) yields SKE up to $\mathcal{O}(\hbar^1)$,
\begin{widetext}
\begin{align}\nonumber\label{CKT_vector_2}
&0=\delta(q^2-m^2)\Bigg[q\cdot\Delta f_V
+\hbar\Big(\frac{E_{\mu}S^{\mu\nu}_{a(n)}}{q\cdot n}\Delta_{\nu} + S^{\mu\nu}_{a(n)}(\partial_{\mu}F_{\rho\nu})\partial^{\rho}_q
+\big(\partial_{\mu}S^{\mu\nu}_{a(n)}\big)\Delta_{\nu}\Big)f_A
\Bigg]
-\frac{\delta'(q^2-m^2)}{q\cdot n}B^{\mu}
\Box_{\mu\nu}\tilde{a}^{\nu} 
\\
&
+\frac{\hbar}{2}\delta(q^2-m^2)\epsilon^{\mu\nu\alpha\beta}\Bigg[
\Delta_{\mu}\left(\frac{n_{\beta}}{q\cdot n}\right)\big[(\Delta_{\nu}a_{\alpha})+F_{\nu\alpha}\big]
+\frac{n_{\beta}}{q\cdot n}\Big( 
(\partial_{\mu}F_{\rho\nu})(\partial_q^{\rho}a_{\alpha})
+ \Big[\big(\Delta_{\nu}a_{\alpha}\big)
-F_{\rho\nu}\big(\partial_q^{\rho}a_{\alpha}\big) \Big]\Delta_{\mu}
\Big)
\Bigg]f_A,
\end{align}
\end{widetext}
where $E_{\mu}=n^{\nu}F_{\mu\nu}$, $B^{\mu}=\frac{1}{2}\epsilon^{\mu\nu\alpha\beta}n_{\nu}F_{\alpha\beta}$, and
\begin{eqnarray}
S^{\mu\nu}_{a(n)}=\frac{\epsilon^{\mu\nu\alpha\beta}a_{\alpha}n_{\beta}}{2q\cdot n}
\end{eqnarray}
is the spin tensor.
When $m=0$ and $a^{\mu}=q^{\mu}$, the second line in Eq.~(\ref{CKT_vector_2}) vanishes and the first line reproduces the CKT in the massless case \cite{Hidaka:2016yjf,Hidaka:2017auj,Huang:2018wdl}. The detailed derivation of Eq.~(\ref{CKT_vector_2}) is shown in Appendix~\ref{sec:Scalar/Axial-VectorKineticEquations}. 

\section{Axial Wigner functions/axial-vector kinetic equation} 
The axial part of Wigner functions is obtained from Eqs.~(\ref{eqA1})-(\ref{eqA3}). 
However, unlike the vector part, Eqs.~(\ref{eqA1}) and (\ref{eqA2}) only lead to the modified dispersion relation 
and do not uniquely fix the MC term. We thus obtain 
\begin{align}\label{axial_WF}
\mathcal{A}_{1}^{\mu}=
2\pi\tilde{F}^{\mu\nu}q_{\nu}\delta'(q^2-m^2)f_V+2\pi\delta(q^2-m^2)H^{\mu},
\end{align}
with an undetermined MC term $ H^\mu $ up to a constraint $\delta(q^2-m^2) q\cdot H=0$ at the on-shell. 
While Eq.~(\ref{eqA3}) yields the AKE, we do not find any quantum correction 
when $ H^\mu=0 $ and $F_{\mu\nu}=0$.

In order to find the MC term for $\mathcal{A}_{1}^{\mu}$, we will implement an alternative method 
by constructing Wigner functions directly through the second quantization of free Dirac fields 
as examined in the massless case \cite{Hidaka:2016yjf}. 
The quantized free Dirac field reads \cite{Peskin} 
\begin{eqnarray}
\psi(x)&=&\int \frac{d^3p}{(2\pi)^3}\frac{1}{\sqrt{2E_{\bf p}}}
\sum_s u^s( p)e^{-ip\cdot x}a^s_{\bf p},
\end{eqnarray}
where $E_{\bf p}=\sqrt{|{\bf p}|^2+m^2}$ and we drop anti-fermions for simplicity. 
We have the annihilation (creation) operators $a^{s(\dagger)}_{\bf p}$ and the wave function 
$ u^s(p) = (\sqrt{p\cdot\sigma}\xi^s , \sqrt{p\cdot\bar{\sigma}}\xi^s ) ^T$ 
with $s$ and $\xi^s$ being spin indices and a two component spinor, respectively~\cite{Peskin}. 
Here, $\sigma^{\mu}$ and $\bar{\sigma}^{\mu}$ are four dimensional Pauli matrices, which satisfy $\sigma^{\mu}\bar{\sigma}^{\nu}+\sigma^{\nu}\bar{\sigma}^{\mu}=\bar\sigma^{\mu}{\sigma}^{\nu}+\bar\sigma^{\nu}{\sigma}^{\mu}=2\eta^{\mu\nu}$, with the Minkowski matrix $\eta^{\mu\nu}$. 

The lessor propagator then takes the form 
\begin{eqnarray}\label{S_2nd_quantization}
&&S^{<}(x,y)= \int \frac{d^3p}{(2\pi)^3}
  \int \frac{d^3p'}{(2\pi)^3}
  \frac{1}{\sqrt{2E_{\bf p}}} \frac{1}{\sqrt{2E_{\bf p'}}}
\\\nonumber
&&\qquad\qquad\times
\sum_{s,s'}u^s(p)\bar{u}^{s'}(p')\langle a^{s'\dagger}_{\bf p'}a^s_{\bf p}\rangle e^{ip_-\cdot X- ip_+\cdot Y},
\end{eqnarray}
where $p^{\mu}_{+}=(p+ p')^{\mu}/2$ and $p^{\mu}_{-}=(p- p')^{\mu}$.
The density operator can be written as $\langle a^{s'\dagger}_{\bf p'}a^s_{\bf p}\rangle=\delta_{ss'}N_V({\bf p}, {\bf p'})+\mathcal{A}_{ss'}({\bf p}, {\bf p'})$,
where $\mathcal{A}_{ss'} ({\bf p}, {\bf p'})\neq 0$ when $s\neq s'$.
We parametrize the results of spin sum as 
$ \sum_s\xi_{s}\xi^{\dagger}_s=n\cdot\sigma=I$ and 
$\sum_{s,s'}\xi_{s}\mathcal{A}_{ss'}\xi^{\dagger}_{s'}=S\cdot\sigma$ such that $S\cdot n=0$. After Wigner transformation, we define
\begin{eqnarray}\nonumber
\tilde{f}_{V}(q,X)&\equiv&  \int\frac{d^3p_-}{(2\pi)^3}N_{V}\left(\bm q+\frac{\bm p_-}{2},\bm q-\frac{\bm p_-}{2}\right) e^{-ip_-\cdot X},
\\
\hat{S}_{\mu}(q,X)&\equiv&  \int\frac{d^3p_-}{(2\pi)^3}S_{\mu}\left(\bm q+\frac{\bm p_-}{2},\bm q-\frac{\bm p_-}{2}\right) e^{-ip_-\cdot X},
\end{eqnarray}
where $\hat{S}_{\mu}(q,X)$ is related to the spin four vector.
Further making the $\hbar$ expansion led by the $p^{\mu}_{-}$ expansion for wave functions in analogous to the derivation in Ref.~\cite{Hidaka:2016yjf} for Weyl fermions, Eq.~(\ref{S_2nd_quantization}) yields $(\mathcal{V}/\mathcal{A})^{\mu}$ in terms of $ \tilde f_{V/A}$ with the explicit forms up to $\mathcal{O}(\hbar^1)$, 
	\begin{eqnarray}\label{free_case_V}
	\mathcal{V}^{\mu}&=&2\pi\delta(q^2-m^2)\Big[q^{\mu}f_V+\hbar \frac{\epsilon^{\mu\nu\alpha\beta}n_{\beta}}{2q\cdot n}\partial_{\nu}(a_{\alpha}f_A)\Big],
	\\
\label	{free_case_A}
	\mathcal{A}^{\mu}&=&2\pi\delta(q^2-m^2)\Big[a^{\mu}f_A+\hbar S^{\mu\nu}_{m(n)}\partial_{\nu}f_V\Big]
.
\end{eqnarray}
Note that we have identified the present parameterizations to the previous ones as 
\begin{eqnarray}\label{rel_f}
&&
f_V=\tilde{f}_V\blue{-}\frac{\hbar S^{\mu\nu}_{m(n)}}{q\cdot n}\partial_{\nu}\hat{S}_{\mu},
\\
\label{rel_a_and_S}
&&
a\cdot n f_A= \hat{S}\cdot q,\quad 
a_{\perp\mu} f_A= \frac{(\hat{S}\cdot q)q_{\perp\mu}}{q\cdot n+m}-m \hat{S}_{\mu},
\end{eqnarray}
where the subscripts $ \perp $ denote the components perpendicular to $ n^\mu  $, i.e., 
$v^{\mu}_{\perp}\equiv v^{\mu}-(v\cdot n)n^{\mu}$ for a vector $v^{\mu}$. 
We also introduced the following tensor:
\begin{eqnarray}\label{spin_tensor_m}
S^{\mu\nu}_{m(n)}=\frac{\epsilon^{\mu\nu\alpha\beta}q_{\alpha}n_{\beta}}{2(q\cdot n+m)}=\frac{\epsilon^{\mu\nu\alpha\beta}q_{\alpha}n_{\beta}}{2a\cdot n}
. 
\end{eqnarray}
One may refer to Appendix~\ref{WignerFunctionsFromDiracWave-functions} for the details of computations.
The $\mathcal{V}^{\mu}_1$ in Eqs.~(\ref{free_case_V}) and (\ref{Y_sol_V1}) 
agree with each other when $F_{\mu\nu}=0$. 
Note that the previous constraint $q\cdot a=q^2-m^2$ is satisfied 
if we take $\hat{S}\cdot q=(q\cdot n+m)f_A$, which implies $-\hat{S}^{\mu}=q_{\perp}^{\mu}f_A/(q\cdot n)$ when $m=0$. 
Thus, $(a\cdot n)/(2q\cdot n)$ is identified with the helicity in the massless limit. 
From (\ref{rel_a_and_S}), one can obtain the second equality in Eq.~(\ref{spin_tensor_m}).

In Eq.~(\ref{free_case_A}), one can read off the MC term 
\footnote{To be more precise, for both positive and negative $q\cdot n$, we should write $2S^{\mu\nu}_{m(n)}=\bar{\epsilon}(q\cdot n)\epsilon^{\mu\nu\alpha\beta}q_{\alpha}n_{\beta}/(|q\cdot n|+m)$, where $\bar{\epsilon}(q\cdot n)$ denotes the sign for $q\cdot n$.} 
\begin{eqnarray}\label{Hmu}
H^{\mu}=S^{\mu\nu}_{m(n)}\Delta_{\nu}f_V
.
\end{eqnarray}
We generalize the derivative operator to include a background field in analogy to the massless case \cite{Hidaka:2016yjf}, 
and find that $\mathcal{A}^{\mu}$ has a symmetric form with $\mathcal{V}^{\mu}$ 
under interchanges $q^{\mu}\leftrightarrow a^{\mu}$ and $f_{V}\leftrightarrow f_A$. 
In Eq.~(\ref{free_case_A}), one could absorb $ H^\mu  $ 
by a redefinition $\bar a^{\mu} f_A \equiv a^\mu f_A + \hbar H^\mu$. 
The freedom of such a redefinition reveals itself as the non-uniqueness of the MC term as we saw when solving the master equations (\ref{eqA1}) and (\ref{eqA2}) for $\mathcal{A}^{\mu}$, 
and could occur in the massive case since $ a^\mu $ is a dynamical variable to be determined by the kinetic theory. 
However, it is crucial to explicitly separate the MC term $ H^\mu $ from $  a^\mu$ 
in order to see a smooth reduction to the CKT 
where $ a^\mu $ is no longer an independent dynamical variable and is enslaved by $ q^\mu $. 
The $ H^\mu $ is also important for including the spin-orbit interaction. Note that $H^{\mu}=0$ when $n^{\mu}=q^{\mu}/m$, which is thus omitted in Refs.~\cite{Weickgenannt:2019dks,Gao:2019znl}. Similar to the case for $f_V$, $a^{\mu}f_A$ also obey the following modified frame transformation,
\begin{eqnarray}\nonumber
&&a^{({n'})\mu}f_A^{(n')}-a^{(n)\mu}f_A^{(n)}
\\
&&=\frac{\hbar\epsilon^{\mu\nu\alpha\beta}}{2}\left(\frac{n_{\beta}}{(q\cdot n+m)}-\frac{n'_{\beta}}{(q\cdot n'+m)}\right)q_{\alpha}\Delta_{\nu}f_V.
\end{eqnarray}

Then, plugging Eqs.~(\ref{LO_WF}) and (\ref{axial_WF}) into Eq.~(\ref{eqA3}) and 
carrying out straightforward arrangements, we derive the AKE as 
\begin{widetext}
\begin{eqnarray}\nonumber\label{CKT_axial_final}
0&=&\delta(q^2-m^2)
\Big(q\cdot\Delta (a^{\mu}f_A)+F^{\nu\mu}a_{\nu}f_A\Big)
+\hbar q^{\mu}\Bigg\{\delta(q^2-m^2)\Bigg[(\partial_{\alpha}S_{m(n)}^{\alpha\nu})\Delta_{\nu}+\frac{S^{\alpha\nu}_{m(n)}E_{\alpha}\Delta_{\nu}}{q\cdot n+m}
+S^{\rho\nu}_{m(n)}(\partial_{\rho}F_{\beta\nu})\partial_q^{\beta}\Bigg]
\\\nonumber
&&
-\delta'(q^2-m^2)\frac{q\cdot B}{q\cdot n+m}q\cdot\Delta\Bigg\}f_V
+\hbar m\Bigg\{\frac{\delta(q^2-m^2)\epsilon^{\mu\nu\alpha\beta}}{2(q\cdot n+m)}\Bigg[m(\partial_{\alpha}n_{\beta})\Delta_{\nu}+(mn_{\beta}+q_{\beta})\Bigg(\frac{\big(E_{\alpha}-\partial_{\alpha}(q\cdot n)\big)}{q\cdot n+m}\Delta_{\nu}
\\
&&-(\partial_{\nu}F_{\rho\alpha})\partial_q^{\rho}\Bigg)\Bigg]
+\delta'(q^2-m^2)\frac{(mn_{\beta}+q_{\beta})\tilde{F}^{\mu\beta}}{q\cdot n+m}q\cdot\Delta\Bigg\} f_V
.
\end{eqnarray}
\end{widetext}
The detailed derivation is shown in Appendix~\ref{sec:Scalar/Axial-VectorKineticEquations}.
Taking the massless limit $m\to0$, one immediately finds that $a^{\mu}=q^{\mu}$ from Eq.~(\ref{CKT_axial_final}) 
and the full equation reduces to the CKT in Ref.~\cite{Hidaka:2017auj} multiplied by $q^{\mu}$, 
which manifests the spin alignment along the momentum. 
In contrast, when $m\neq 0$, the background field and the derivative of local frame vector engender nontrivial spin force. 


When solving kinetic equations (\ref{CKT_vector_2}) and (\ref{CKT_axial_final}), 
we need to handle the terms proportional to $\delta'(q^2-m^2)$. 
All these terms can be arranged with the LO kinetic theory shown below Eq.~(\ref{LO_WF}) 
$2\delta'(q^2-m^2)q^\mu q \cdot \Delta f_{V}=-\delta(q^2-m^2) \partial_q^{\mu}(q\cdot\Delta f_{V})$ 
and $2\delta'(q^2-m^2)\Box_{\mu\nu}\tilde{a}^{\nu}=-\delta(q^2-m^2)\partial^{\rho}_{q}\big((q\cdot n)^{-1}n_{\rho}\Box_{\mu\nu}\tilde{a}^{\nu}\big)$ up to $\mathcal{O}(\hbar^1)$. 
Then, all the delta functions can be factored out from the CKTs.

From the solutions of CKTs, one can get 
the vector/axial currents and the symmetric/antisymmetric parts of the canonical energy-momentum tensor~\footnote{
More precisely, one has to include both fermions and anti-fermions, 
and put an overall sign function $\bar{\epsilon}(q\cdot n)$ in front of the Wigner function.},
\begin{align}\label{def_J_Tmunu}
J^{\mu}_{V/5}=4\int_q(\mathcal{V}/\mathcal{A})^{\mu}, 
\quad
T^{\mu\nu}_{S/A}=2\int_q\big(\mathcal{V}^{\mu}q^{\nu}\pm \mathcal{V}^{\nu}q^{\mu}\big),
\end{align} 
where $\int_q\equiv\int d^4q/(2\pi)^4$. 
Angular-momentum conservation arises from Eq.~(\ref{eqv3}) as discussed in the massless case \cite{Yang:2018lew}, 
and $T^{\mu\nu}_A$ 
is responsible for angular-momentum transfer (see Ref.~\cite{Hattori:2019lfp} and references therein). 
As an example, we consider the non-relativistic limit with constant $n^{\mu}$ and $E^{\mu}$. By approximating $q^{\mu}\approx mn^{\mu}$, Eq.~(\ref{CKT_axial_final}) yields $n\cdot\Delta \big(a^{\mu}f_A-\hbar\epsilon^{\mu\nu\alpha\beta}E_{\alpha}n_{\beta}\partial_{q\nu}(f_V/4)\big)\approx 0$ 
after dropping the sub-leading terms in $m$ 
and arranging the delta functions with the aforementioned strategy. 
Then, we find a spin-Hall current in the stationary state 
\begin{eqnarray}
J^{\mu}_5\approx-2\pi\hbar\epsilon^{\mu\nu\alpha\beta}E_{\alpha}n_{\beta}\int_q\delta(q^2-m^2)\partial_{q\nu}f_V
.
\end{eqnarray}

%

\section{Anomalous transport in thermal equilibrium} 
As an application, we discuss the mass effects on the anomalous transport in global equilibrium with constant thermal vorticity and chemical potentials, 
and compare our conclusions from the SKE and AKE with those from the Kubo formula calculations.

While collisionless kinetic equations do not uniquely determine equilibrium WFs \cite{Hidaka:2017auj}, 
we may construct equilibrium WFs motivated by the following considerations. 
For the vector charges, we may naturally take the Fermi distribution function 
$f_{V\text{eq}}=f_0(q\cdot u-\mu_V)=1/\big(\exp(\beta (q\cdot u-\mu_V))+1\big)$, 
where $\beta=1/T$ and $\mu_V$ are the inverse temperature and vector chemical potential, respectively. 
On the other hand, the axial charge should be damped out as $t\rightarrow \infty$ when $m\neq 0$ 
because of the scattering.
Thus, $f_{A\text{eq}}$ may be at most $\mathcal{O}(\hbar^1)$ 
induced by the vorticity correction. 
Referring to the massless case \cite{Chen:2015gta,Hidaka:2017auj}, 
we also expect that $\mathcal{A}^{\mu}_{\text{eq}}$ does not have an explicit dependence on $n^{\mu}$. 
Thus we propose an equilibrium Wigner function in constant magnetic field and thermal vorticity
\begin{eqnarray}\label{equ_V_mu}
\mathcal{V}^{\mu}_{\text{eq}}&=&2\pi\delta(q^2-m^2)q^{\mu}f_0,
\\\nonumber
\mathcal{A}^{\mu}_{\text{eq}}&=&2\pi\hbar\Big[\frac{\delta(q^2-m^2)}{4}q_{\nu}\epsilon^{\nu\mu\alpha\beta}\Omega_{\alpha\beta}\partial_{q\cdot \beta}
\\
&&+\tilde{F}^{\mu\nu}q_{\nu}\delta'(q^2-m^2)\Big]f_0,
\label{equ_A_mu}
\end{eqnarray} 
where $\Omega_{\mu\nu}=\partial_{[\mu}(\beta_{\nu]})/2$ corresponds to the thermal vorticity and  $\beta_{\nu}=\beta u_{\nu}$. 
The equilibrium $\mathcal{A}^{\mu}_{\text{eq}}$ takes the equivalent form 
as the one for massless fermions at constant temperature except for the on-shell condition \cite{Hidaka:2017auj}, 
and was also proposed for massive fermion \cite{Becattini2013a} (similar form in \cite{Fang:2016vpj}) which satisfies the master equations. See Refs.\cite{Prokhorov:2017atp, Prokhorov:2018qhq} for WFs beyond weak vorticity and with acceleration.

The equilibrium Wigner functions (\ref{equ_V_mu}) and (\ref{equ_A_mu}) 
now lead to CSE and axial-CVE, $J^{\mu}_{B/\omega 5}=\sigma_{B/\omega}(B/\omega)^{\mu}$, where
\begin{eqnarray}
\sigma_{B/\omega}=\frac{\hbar}{2\pi^2}\int^{\infty}_0d|{\bf q}|g_{B/\omega}f^{(-/+)}_{0}(E_{\bf q})
\end{eqnarray} 
with $g_{B}=1$, $g_{\omega}=(2E_{\bf q}^2-m^2)/E_{\bf q}$, and $f^{(\pm)}_{0}=f_{0}(E_{\bf q}-\mu_{V})\pm f_{0}(E_{\bf q}+\mu_{V})$. Here the fluid vorticity $\omega^{\mu}$ is defined as $\omega^{\mu}\equiv Tu_{\nu}\epsilon^{\mu\nu\alpha\beta}\Omega_{\alpha\beta}/2=\epsilon^{\mu\nu\alpha\beta}u_{\nu}(\partial_{\alpha}u_{\beta})/2$.
The above results agree with those derived from the Kubo formula with thermal correlators \cite{Buzzegoli:2017cqy,Lin:2018aon}. 
Similar to the massless case \cite{Chen:2014cla,Chen:2015gta,Hidaka:2017auj}, 
a part of the axial-CVE comes from the MC term 
which can be identified by comparing $\mathcal{A}^{\mu}_{\text{eq}}$ 
with the general form (\ref{free_case_A}).

Finally, since $f_{A\text{eq}}=\mathcal{O}(\hbar^1)$, we conclude that 
the CME and vector-CVE vanish in an equilibrium when $m\neq 0$. 
On the other hand, it was shown by the thermal field theoretical calculation 
that the CME in an equilibrium receives no mass correction~\cite{Fukushima:2009ft, Fukushima:2012vr}. 
However, the axial chemical potential $ \mu_5 $ is not a static quantity in the massive case, 
and the thermal field theory with a constant $ \mu_5 $ does not correctly capture its dynamics. 
The thermal field theoretical calculation may work only under certain caveats on the existence of $ \mu_5 $, 
and there are no equilibrium currents in a strict thermal equilibrium at $ \mu_5 = 0 $. Only when equilibrium statistical operators breaking charge conjugation, parity, and rotation symmetry (e.g., with acceleration and chemical potential) exist, vector currents would be allowed.

\section{Conclusions and outlook}
In this work, we developed the quantum kinetic theory for arbitrary-mass fermions, 
which provides a theoretical framework for describing the coupled dynamics among spin and the vector and axial charges. 
Moreover, we have constructed a bridge on the longstanding gap between the CKT and axial kinetic theory. 
In the future, we will include collision effects to investigate their relaxation dynamics. 
It is feasible with an extension of the collision terms 
developed in the massless limit~\cite{Hidaka:2016yjf,Hidaka:2017auj,Hidaka:2018ekt}, 
and with deeper understandings and techniques obtained in this work. 

 
\acknowledgments
The authors thank Igor Shovkovy and Qun Wang for fruitful discussions during the Workshop 
on Recent Developments in Chiral Matter and Topology, 
and Xu-Guang Huang, Yu-Chen Liu, and Kazuya Mameda when K. H. visited Fudan University.
Y. H. was partially supported by Japan Society of Promotion of Science (JSPS), Grants-in-Aid for Scientific Research(KAKENHI) Grants No. 15H03652, 16K17716, and 17H06462.  Y. H. was also partially supported by RIKEN iTHES Project and iTHEMS Program. K. H. and D.-L. Y. are supported in part by Yukawa International Program for Quark-hadron Sciences (YIPQS). D.-L. Y. is also partially supported by Keio Institute of Pure and Applied Sciences (KiPAS) project in Keio University.

\appendix
\section{Derivation of Master Equations}\label{sec:DerivationOfMasterEquations}
In this section, we derive the six master equations for Wigner functions of Dirac fermions. 
We shall start from the Dirac Lagrangian density,
\begin{eqnarray}
	\mathscr{L}=\bar{\psi}(i\slashed{\mathcal{D}}-m)\psi,
\end{eqnarray}
where the covariant derivative is $\mathcal{D}_{\mu}=\partial_{\mu}+iA_{\mu}/\hbar$ with the U(1) gauge field $A_{\mu}$.
We define the greater and lessor propagators as
\begin{align}
	[S^{>}(x,y)]_{\alpha\beta}&\equiv e^{\frac{i}{\hbar}\int_{x}^{y} A_{\mu}(z)dz^{\mu}}\langle \psi_{\alpha}(x)\bar{\psi}_{\beta}(y)\rangle,\\
	[S^{<}(x,y)]_{\alpha\beta}&\equiv e^{\frac{i}{\hbar}\int_{x}^{y} A_{\mu}(z)dz^{\mu}}\langle \bar{\psi}_{\beta}(y)\psi_{\alpha}(x)\rangle .
\end{align}
Here, $\alpha$ and $\beta$ denote the spinor indices.

After the Wigner transformation defined as
\begin{equation}
	\begin{split}
		S^{\lessgtr}(q,X) = \int \frac{d^{4}Y}{(2\pi)^{4}}e^{\frac{i}{\hbar}q\cdot Y} S^{\lessgtr}(X+Y/2,X-Y/2),
	\end{split}
\end{equation}
the lessor propagator without collision terms obeys 
\begin{eqnarray}\label{KB_eq0}\nonumber
&&(\slashed{\Pi}-m)S^{<}+\gamma^{\mu}i\frac{\hbar}{2}\nabla_{\mu}S^{<}=0,
\\
&&S^{<}(\slashed{\Pi}-m)-i\frac{\hbar}{2}\nabla_{\mu}S^{<}\gamma^{\mu}=0,
\end{eqnarray}
or equivalently,
\begin{eqnarray}\nonumber\label{KB_eq}
	&&\{(\slashed{\Pi}-m), S^< \} +\frac{i\hbar}{2}[\gamma^{\mu}, \nabla_{\mu}S^{<}]=0,
\\
	&&[(\slashed{\Pi}-m), S^{<}]+\frac{i\hbar}{2}\{\gamma^{\mu}, \nabla_{\mu}S^{<}\}
	=0,
\end{eqnarray}
where we introduce
\begin{eqnarray}
	\nonumber
	&&\nabla_{\mu}=\partial_{\mu}+j_0(\Box)F_{\nu\mu}\partial^{\nu}_{q},
	\\ &&\Pi_{\mu}=q_{\mu}+\frac{\hbar}{2}j_1(\Box)F_{\nu\mu}\partial^{\nu}_{q},\quad\Box=\frac{\hbar}{2}\partial_{\rho}\partial^{\rho}_q.
\end{eqnarray} 
Here, we define $\partial_{\mu}\equiv\partial/\partial X^{\mu}$,   $\partial_{q}^{\mu}\equiv\partial/\partial q_{\mu}$, and $F_{\mu\nu}\equiv \partial_{\mu}A_{\nu}-\partial_{\nu}A_{\mu}$. Also, $j_0(\Box),j_1(\Box)$ are modified Bessel functions. We note that $\partial_{\rho}$ in $\Box$ only act on $F_{\nu\mu}$ when having spacetime-dependent background fields. 
Making the $\hbar$ expansion, which corresponds to the gradient expansion for $\partial_{\mu}\ll q_{\mu}$, one finds
\begin{eqnarray}\nonumber
	\nabla_{\mu}&=&\partial_{\mu}+F_{\nu\mu}\partial^{\nu}_q-\frac{\hbar^2}{24}(\partial_{\rho}\partial_{\lambda}F_{\nu\mu})\partial^{\lambda}_q\partial^{\rho}_q\partial^{\nu}_{q}+\mathcal{O}(\hbar^4), 
	\\
	\Pi_{\mu}&=&q_{\mu}+\frac{\hbar^2}{12}(\partial_{\rho}F_{\nu\mu})\partial^{\rho}_q\partial^{\nu}_{q}+\mathcal{O}(\hbar^4).
\end{eqnarray}
We then apply the decomposition based on the Clifford algebra,
\begin{eqnarray}
	S^<=\mathcal{S}+ i\mathcal{P}\gamma^5+ \mathcal{V}_{\mu}\gamma^\mu+\mathcal{A}_{\mu}\gamma^5\gamma^{\mu}+ \frac{\mathcal{S}_{\mu\nu}}{2}\Sigma^{\mu\nu},
\end{eqnarray}
where $\Sigma^{\mu\nu}=i[\gamma^{\mu},\gamma^{\nu}]/2$ and $\gamma^5=i\gamma^0\gamma^1\gamma^2\gamma^3$.

For simplicity, we consider the collisionless case, in which Eq.~(\ref{KB_eq}) result in
\begin{align}\label{meq_1}
	m\mathcal{V}_{\mu}&=\Pi_{\mu}\mathcal{S}-\frac{\hbar}{2}\nabla^{\nu}\mathcal{S}_{\nu\mu},
	\\
	2m\mathcal{A}^{\mu}&=-\epsilon^{\mu\nu\rho\sigma}\Pi_{\sigma}\mathcal{S}_{\nu\rho}
	+\hbar\nabla^{\mu}\mathcal{P},
	\\\label{meq_3}
	m\mathcal{S}&=\Pi\cdot\mathcal{V},
	\\\label{meq_4}
	m\mathcal{P}&=- \frac{\hbar}{2}\nabla_{\mu}\mathcal{A}^{\mu},
	\\\label{meq_5}
	m\mathcal{S}_{\mu\nu}&=-\epsilon_{\mu\nu\rho\sigma} \Pi^{\rho}\mathcal{A}^{\sigma}
	+\frac{\hbar}{2}\nabla_{[\mu}\mathcal{V}_{\nu]}
	,
\end{align}
and
\begin{align}
	\nabla^{\mu}\mathcal{V}_{\mu}&=0,\\
	\Pi\cdot\mathcal{A}&=0,\\
	\Pi_{\nu}\mathcal{V}_{\mu}-\Pi_{\mu}\mathcal{V}_{\nu}
	-\frac{1}{2}\epsilon_{\mu\nu\rho\sigma} \hbar\nabla^{\rho}\mathcal{A}^{\sigma}
	&=0,\\
	2\Pi^{\nu} \mathcal{S}_{\nu\mu}+ \hbar\nabla_{\mu}\mathcal{S} &=0,
	\\\label{last_eq}
	\epsilon^{\mu\nu\rho\sigma}\hbar\nabla_{\sigma}\mathcal{S}_{\nu\rho}+4\Pi^{\mu}\mathcal{P}&=0.
\end{align}

By writing $\mathcal{S}$, $\mathcal{P}$, and $\mathcal{S}_{\mu\nu}$ in terms of $\mathcal{V}_{\mu}$ and $\mathcal{A}_{\mu}$ from Eqs.~(\ref{meq_3})-(\ref{meq_5}), we obtain
\begin{eqnarray}\label{rrr1}
	&&\nabla\cdot\mathcal{V}=0,
	\\\label{rrr2}\nonumber
	&&(\Pi_{\mu}\Pi\cdot\mathcal{V}-m^2\mathcal{V}_{\mu})
	\\
	&&=-\frac{\hbar}{2}\epsilon^{\nu\mu\rho\sigma}\nabla^{\nu}\Pi^{\rho}\mathcal{A}^{\sigma}+\frac{\hbar^2}{4}\nabla^{\nu}\nabla_{[\nu}\mathcal{V}_{\mu]},
	,
	\\\label{rrr3}
	&&\Pi\cdot\mathcal{A}=0,
	\\\label{rrr4}
	&&\Pi_{\nu}\mathcal{V}_{\mu}-\Pi_{\mu}\mathcal{V}_{\nu}
	=\frac{\hbar}{2}\epsilon_{\mu\nu\rho\sigma}\nabla^{\rho}\mathcal{A}^{\sigma},
	\\\label{rrr5}\nonumber
	&&(\Pi^2-m^2)\mathcal{A}^{\mu}-\Pi_{\sigma}\Pi^{\mu}\mathcal{A}^{\sigma}
	\\
	&&=\frac{\hbar}{2}\epsilon^{\mu\nu\rho\sigma}\Pi_{\sigma}
	\nabla_{\nu}\mathcal{V}_{\rho}-\frac{\hbar^2}{4}\nabla^{\mu}\nabla\cdot\mathcal{A}
	,
	\\\label{rrr6}
	&&\hbar\big((\Pi^{\mu}\nabla_{\sigma})\mathcal{A}^{\sigma}+\nabla_{\sigma}\Pi^{\sigma}\mathcal{A}^{\mu}\big)
	=\frac{\hbar^2}{4}\epsilon^{\mu\nu\rho\sigma}\nabla_{\sigma}\nabla_{[\nu}\mathcal{V}_{\rho]}
	,
	\\\label{rrr7}
	&&\hbar\big(\Pi\cdot\nabla\mathcal{V}_{\mu}
	+(\nabla_{\mu}\Pi^{\nu})\mathcal{V}_{\nu}\big)
	=2\epsilon_{\nu\mu\rho\sigma}(\Pi^{\nu}\Pi^{\rho})\mathcal{A}^{\sigma},
\end{eqnarray}
where $A_{[\mu}B_{\nu]}\equiv A_{\mu}B_{\nu}-B_{\nu}A_{\mu}$. 
Up to $\mathcal{O}(\hbar)$, Eqs.~(\ref{rrr1})-(\ref{rrr7}) read
\begin{eqnarray}\label{nr1}
	\Delta\cdot\mathcal{V}&=&0,
	\\\label{nr2}
	q_{\mu}q\cdot\mathcal{V}-m^2\mathcal{V}_{\mu}&=&-\hbar\tilde{F}_{\mu\sigma}\mathcal{A}^{\sigma}+\frac{\hbar}{2}\epsilon_{\mu\nu\rho\sigma}q^{\rho}\Delta^{\nu}\mathcal{A}^{\sigma}
	,
	\\\label{nr3}
	q\cdot\mathcal{A}&=&0,
	\\\label{nr4}
	q_{\nu}\mathcal{V}_{\mu}-q_{\mu}\mathcal{V}_{\nu}
	&=&\frac{\hbar}{2}\epsilon_{\mu\nu\rho\sigma}\Delta^{\rho}\mathcal{A}^{\sigma},
	\\\label{nr5}
	(q^2-m^2)\mathcal{A}^{\mu}&=&\frac{\hbar}{2}\epsilon^{\mu\nu\rho\sigma}q_{\sigma}
	\Delta_{\nu}\mathcal{V}_{\rho}
	,
	\\\label{nr6}
	q\cdot\Delta\mathcal{A}^{\mu}+F^{\nu\mu}\mathcal{A}_{\nu}
	&=&\frac{\hbar}{2}\epsilon^{\mu\nu\rho\sigma}\Delta_{\sigma}\Delta_{\nu}\mathcal{V}_{\rho}
	\nonumber
	\\
	&=&\frac{\hbar}{2}\epsilon^{\mu\nu\rho\sigma}(\partial_{\sigma}F_{\beta\nu})\partial^{\beta}_{q}\mathcal{V}_{\rho}
	,
	\\\label{nr7}
	q\cdot\Delta \mathcal{V}^{\mu}+F^{\nu\mu}\mathcal{V}_{\nu}&=&\frac{\hbar}{2}\epsilon_{\mu\nu\rho\sigma}(\partial^{\nu}F^{\beta\rho})\partial_{q\beta}\mathcal{A}^{\sigma},
\end{eqnarray}
where $\Delta_{\mu}=\partial_{\mu}+F_{\nu\mu}\partial_q^{\nu}$.
To obtain the right-hand side of Eq.~(\ref{nr7}), we employ the following equation,
\begin{eqnarray}\nonumber
	2\epsilon_{\nu\mu\rho\sigma}(\Pi^{\nu}\Pi^{\rho})\mathcal{A}^{\sigma}&=&\frac{1}{6}\epsilon_{\nu\mu\rho\sigma}\big(\partial^{\rho}F^{\beta\nu}+\partial^{\beta}F^{\rho\nu}\big)\partial_{q\beta}\mathcal{A}^{\sigma}
	\\
	&=&\frac{1}{2}\epsilon_{\nu\mu\rho\sigma}(\partial^{\rho}F^{\beta\nu})\partial_{q\beta}\mathcal{A}^{\sigma},
\end{eqnarray}
where we derive the second equality above from the relation 
\begin{eqnarray}
	\epsilon_{\nu\mu\rho\sigma}\partial^{\beta}F^{\rho\nu}=2\epsilon_{\nu\mu\rho\sigma}\partial^{\rho}F^{\beta\nu}
\end{eqnarray}
led by
\begin{eqnarray}\nonumber
	\epsilon_{\nu\mu\rho\sigma}\partial^{\beta}F^{\rho\nu}&=&\epsilon_{\nu\mu\rho\sigma}\partial^{\beta}(\partial^{\rho}A^{\nu}-\partial^{\nu}A^{\rho})
	\\\nonumber
	&=&\epsilon_{\nu\mu\rho\sigma}\Big(\partial^{\rho}(F^{\beta\nu}+\partial^{\nu}A^{\beta})-\partial^{\beta}\partial^{\nu}A^{\rho}\Big)
	\\
	&=&\epsilon_{\nu\mu\rho\sigma}\Big(\partial^{\rho}F^{\beta\nu}+\frac{1}{2}\partial^{\beta}F^{\rho\nu}\Big).
\end{eqnarray}
Note that Eq.~(\ref{nr7}) is in fact redundant, which can be derived from Eq.~(\ref{nr4}). On the other hand, one can further rewrite Eq.~(\ref{nr2}) by using Eq.~(\ref{nr4}) into a form similar to Eq.~(\ref{nr5}). Accordingly, we shall only deal with the following six master equations,
\begin{eqnarray}\label{Aeqv1}
	\Delta\cdot\mathcal{V}&=&0,
	\\\label{Aeqv2}
	(q^2-m^2)\mathcal{V}_{\mu}&=&-\hbar\tilde{F}_{\mu\nu}\mathcal{A}^{\nu}_,
	\\\label{Aeqv3}
	q_{\nu}\mathcal{V}_{\mu}-q_{\mu}\mathcal{V}_{\nu}
	&=&\frac{\hbar}{2}\epsilon_{\mu\nu\rho\sigma}\Delta^{\rho}\mathcal{A}^{\sigma},
	\\\label{AeqA1}
	q\cdot\mathcal{A}&=&0,
	\\\label{AeqA2}
	(q^2-m^2)\mathcal{A}^{\mu}&=&\frac{\hbar}{2}\epsilon^{\mu\nu\rho\sigma}q_{\sigma}
	\Delta_{\nu}\mathcal{V}_{\rho}
	,
	\\\label{AeqA3}\nonumber
	q\cdot\Delta\mathcal{A}^{\mu}+F^{\nu\mu}\mathcal{A}_{\nu}
	&=&
	\frac{\hbar}{2}\epsilon^{\mu\nu\rho\sigma}(\partial_{\sigma}F_{\beta\nu})\partial^{\beta}_{q}\mathcal{V}_{\rho}
	\\
	&=&\frac{\hbar}{2}(\partial_{\alpha}\tilde{F}^{\mu\nu})\partial^{\alpha}_q\mathcal{V}_{\nu}
	,
\end{eqnarray}
where $\tilde{F}^{\mu\nu}=\epsilon^{\mu\nu\alpha\beta}F_{\alpha\beta}/2$ and we employ the Schouten identity,
\begin{eqnarray}\label{Schouten}\nonumber
	\eta^{\lambda}_{\mu}\epsilon_{\rho\nu\alpha\beta}-\eta^{\lambda}_{\rho}\epsilon_{\mu\nu\alpha\beta}-\eta^{\lambda}_{\nu}\epsilon_{\rho\mu\alpha\beta}-\eta^{\lambda}_{\alpha}\epsilon_{\rho\nu\mu\beta}-\eta^{\lambda}_{\beta}\epsilon_{\rho\nu\alpha\mu}=0,
	\\
\end{eqnarray}
to derive the last equality in Eq.~(\ref{AeqA3}). 

\section{Perturbative Solution for Wigner Functions}\label{eq:PerturbativeSolutionForWignerFunctions}
We will then seek for the perturbative solution for $\mathcal{V}^{\mu}$ and $\mathcal{A}^{\mu}$ from the equations above, for which we take $(\mathcal{V}/\mathcal{A})_{\mu}=(\mathcal{V}/\mathcal{A})_{0\mu}+\hbar (\mathcal{V}/\mathcal{A})_{1\mu}+\mathcal{O}(\hbar^2)$.
At the leading order up to $\mathcal{O}(1)$, from Eqs.~(\ref{Aeqv2}) and (\ref{Aeqv3}), it is found 
\begin{eqnarray}
	\mathcal{V}_{0\mu}=2\pi q_{\mu}\delta(q^2-m^2)f_V,
\end{eqnarray}
which follows the leading-order kinetic theory led by Eq.~(\ref{Aeqv1}),
\begin{eqnarray}\label{LO_V}
	\delta(q^2-m^2)q\cdot\Delta f_V=\mathcal{O}(\hbar).
\end{eqnarray}
The $2\pi$ factor in Eq.~(\ref{LO_V}) is introduced for convention.
For the axial part, Eqs.~(\ref{AeqA1}) and (\ref{AeqA2}) yield
\begin{eqnarray}
	\mathcal{A}^{\mu}_0=2\pi a^{\mu}\delta(q^2-m^2)f_A,  
\end{eqnarray}
where 
\begin{eqnarray}\label{LO_A}
	a\cdot q=q^2-m^2
\end{eqnarray}
satisfies $q\cdot \mathcal{A}=0$ with the onshell condition. Now, plugging Eq.~(\ref{LO_A}) into Eq.~(\ref{AeqA3}), we find
\begin{eqnarray}
	\delta(q^2-m^2)
	\Big(q\cdot\Delta (a^{\mu}f_A)+F^{\nu\mu}a_{\nu}f_A\Big)=\mathcal{O}(\hbar),
\end{eqnarray}
which corresponds to the Bargmann-Michel-Telegdi (BMT) equation.

Subsequently, according to Eqs.~(\ref{Aeqv1})-(\ref{AeqA3}), for the next-to-leading-order solution up to $\mathcal{O}(\hbar)$, we then have to solve 
\begin{eqnarray}\label{rr1}
	&&\Delta\cdot\mathcal{V}=0,
	\\\label{rr2}
	&&(q^2-m^2)\mathcal{V}_{1\mu}=-\tilde{F}_{\mu\nu}\mathcal{A}^{\nu}_{0},
	\\\label{rr3}
	&&q_{\nu}\mathcal{V}_{1\mu}-q_{\mu}\mathcal{V}_{1\nu}
	=\frac{1}{2}\epsilon_{\mu\nu\rho\sigma}\Delta^{\rho}\mathcal{A}^{\sigma}_{0},
	\\\label{rr4}
	&&q\cdot\mathcal{A}=0,
	\\\label{rr5}
	&&(q^2-m^2)\mathcal{A}^{\mu}_1=\frac{1}{2}\epsilon^{\mu\nu\rho\sigma}q_{\sigma}
	\Delta_{\nu}\mathcal{V}_{0\rho}
	,
	\\\label{rr6}\nonumber
	&&q\cdot\Delta\mathcal{A}_1^{\mu}+F^{\nu\mu}\mathcal{A}_{1\nu}
	=
	\frac{1}{2}\epsilon^{\mu\nu\rho\sigma}(\partial_{\sigma}F_{\beta\nu})\partial^{\beta}_{q}\mathcal{V}_{0\rho}
	\\
	&&
	=\frac{1}{2}(\partial_{\alpha}\tilde{F}^{\mu\nu})\partial^{\alpha}_q\mathcal{V}_{0\nu}
	.
\end{eqnarray}

The vector part $\mathcal{V}_1$ can be solved from Eqs.~(\ref{rr2}) and (\ref{rr3}) in analogous to the massless case. Here Eq.~(\ref{rr3}) follows the same structure as the massless master equation to solve for the side-jump term. It is found
\begin{eqnarray}\label{AY_sol_V1}\nonumber
	\mathcal{V}_{\mu}&=&2\pi\delta(q^2-m^2)q_{\mu}f_V+2\pi\hbar\tilde{F}_{\mu\nu}a^{\nu}\delta'(q^2-m^2)f_A
	\\
	&&+2\pi\hbar \delta(q^{2}-m^{2})G_{\mu},
\end{eqnarray} 
where 
\begin{widetext}
\begin{eqnarray}\nonumber
\delta(q^{2}-m^{2})G_{\mu}
&=&\frac{\delta(q^{2}-m^{2})}{2n\cdot q}\epsilon_{\mu\nu\rho\sigma}n^{\nu} \Delta^{\rho}(a^{\sigma}f_A)-\frac{\delta'(q^{2}-m^{2})}{n\cdot q}\tilde{F}_{\mu\nu}n^{\nu}q_{\sigma}(a^{\sigma}f_A)
\nonumber
\\
&=&\frac{\delta(q^{2}-m^{2})}{2n\cdot q}\Big(\epsilon_{\mu\nu\rho\sigma}n^{\nu} (\Delta^{\rho}a^{\sigma}f_A)+2\tilde{F}_{\mu\nu}n^{\nu}f_A\Big). 
	\label{eq:G}
\end{eqnarray}
\end{widetext}
Here $n^{\mu}$ corresponds to a frame vector in the analogous to the massless case.
We employed the relation $(q^{2}-m^{2})\delta'(q^{2}-m^{2})=-\delta(q^{2}-m^{2})$ to obtain the last line of Eq.~\eqref{eq:G}.
We will later utilize the solution in Eq.~(\ref{AY_sol_V1}) to derive the scalar kinetic theory from Eq.~(\ref{rr1}). Since $G_{\mu}$ reduces to the side-jump term when $m=0$ and contributes to the magnetization currents, we will call $G_{\mu}$ as the magnetization-current (MC) term. 

For the axial part $\mathcal{A}_{1\mu}$, from Eqs.~(\ref{rr4}) and (\ref{rr5}), it is found
\begin{eqnarray}\nonumber
	\mathcal{A}_{\mu}&=&2\pi\delta(q^2-m^2) a_{\mu}f_A
	+\hbar\tilde{F}_{\mu\nu}q^{\nu}2\pi\delta'(q^2-m^2)f_V
	\\
	&&+2\pi\hbar \delta(q^2-m^2)H_{\mu},
\end{eqnarray}
where $q\cdot H=0$. Based on the side-jump term in the massless limit, it is expected that the MC term here reads
\begin{eqnarray}
	H^{\mu}=g\epsilon^{\mu\nu\alpha\beta}q^{\alpha}n^{\beta}\Delta_{\nu}f_V,
\end{eqnarray}
where $g\rightarrow 1/(2q\cdot n)$ when $m\rightarrow 0$. However, given that we are unable to fix $g$ by Eqs.~(\ref{rr4}) and (\ref{rr5}). We shall implement an alternative way to derive $g$ from the free WFs in the absence of background fields obtained from the free Dirac fields in the following section. According to Eq.~(\ref{free_case}), we find $2g=\text{sgn}(q\cdot n)/(|q\cdot n|+m)$ assuming it remains unchanged in the presence of background fields.

\section{Wigner functions from Dirac wave-functions}\label{WignerFunctionsFromDiracWave-functions}
In this section, we employ an alternative method to derive WFs without background fields up to $\mathcal{O}(\hbar)$. In particular, we will utilize the result to determine the MC term in $\mathcal{A}_{\mu}$.
We will start from the 2nd quantization of the free Dirac fields~\cite{Peskin},
\begin{eqnarray}\nonumber
	\psi(x)&=&\int \frac{d^3p}{(2\pi)^3}\frac{1}{\sqrt{2E_{\bf p}}}
	\sum_s\left(u^s( p)e^{-ip\cdot x}a^s_{\bf p}+v^s(p)e^{ip\cdot x}b^{s\dagger}_{\bf p}\right),
	\\\nonumber
	\bar{\psi}(x)&=&\int \frac{d^3p}{(2\pi)^3}\frac{1}{\sqrt{2E_{\bf p}}}\sum_s\left(\bar{u}^s(p)e^{ip\cdot x}a^{s\dagger}_{\bf p}+\bar{v}^s( p)e^{-ip\cdot x}b^s_{\bf p}\right), 
	\\
\end{eqnarray}
where
\begin{eqnarray}
	u^s(p) = 
	\begin{pmatrix}
		\sqrt{p\cdot\sigma}\xi^s 
		\\
		\sqrt{p\cdot\bar{\sigma}}\xi^s 
	\end{pmatrix},
	\quad
	v^s(p) = 
	\begin{pmatrix}
		\sqrt{p\cdot\sigma}\eta^s
		\\
		-\sqrt{p\cdot\bar{\sigma}}\eta^s 
	\end{pmatrix}.
\end{eqnarray}
Here, $\sigma^{\mu}$ and $\bar{\sigma}^{\mu}$ are four dimensional Pauli matrices, which satisfy $\sigma^{\mu}\bar{\sigma}^{\nu}+\sigma^{\nu}\bar{\sigma}^{\mu}=\bar\sigma^{\mu}{\sigma}^{\nu}+\bar\sigma^{\nu}{\sigma}^{\mu}=2\eta^{\mu\nu}$, with the Minkowski matrix $\eta^{\mu\nu}$.
For simplicity, we will drop the anti-fermions,
\begin{widetext}
\begin{eqnarray}
	S^{<}(x,y)=\langle\bar{\psi}(y)\psi(x)\rangle
	=
	\int \frac{d^3p}{(2\pi)^3\sqrt{2E_{\bf p}}}\int \frac{d^3p'}{(2\pi)^3\sqrt{2E_{\bf p'}}}\sum_{s,s'}\left(u^s(p)\bar{u}^{s'}(p')\langle a^{s'\dagger}_{\bf p'}a^s_{\bf p}\rangle e^{i(p'-p)\cdot X-\frac{i}{2}(p'+p)\cdot Y}
	\right).
\end{eqnarray}
\end{widetext}
Here we make change of coordinates by taking $X=(x+y)/2$ and $Y=x-y$ in the second equality. We then carry out the Wigner transformation, which yields
\begin{widetext}
\begin{eqnarray}\nonumber
	&&\int d^4Y e^{iq\cdot Y}S^<(x,y)
	\\\nonumber
	&&=\int\frac{d^3p_-}{(2\pi)^3}\int\frac{d^3p_+}{(2\pi)^3}(2\pi)^4e^{-ip_-\cdot X}\delta^4\left(q-p_+\right)
	\sum_{s,s'}u^s\left(p_{+}+\frac{p_-}{2}\right)\bar{u}^{s'}\left(p_{+}-\frac{p_-}{2}\right)\Big\langle a^{s'\dagger}_{\bm p_{+}-\frac{\bm p_-}{2}}a^s_{\bm p_{+}+\frac{\bm p_-}{2}}\Big\rangle
	\\
	&&=\pi\int\frac{d^3p_-}{(2\pi)^3}\frac{e^{-ip_-\cdot X}\delta\left(q_0-p_{+0}\right)}{\left(\left(|{\bf q}|^2+\frac{|\bf p_-|^2}{4}+m^2\right)^2-({\bf p_-\cdot q})^2\right)^{1/4}}\sum_{s,s'}u^s\left(p_{+}+\frac{p_-}{2}\right)\bar{u}^{s'}\left(p_{+}-\frac{p_-}{2}\right)\Big\langle a^{s'\dagger}_{\bm p_{+}-\frac{\bm p_-}{2}}a^s_{\bm p_{+}+\frac{\bm p_-}{2}}\Big\rangle,
\end{eqnarray}
\end{widetext}
where $p_+=(p+p')/2$ and $p_-=p-p'$. We now define the density operators as
\begin{eqnarray}
	\langle a^{s'\dagger}_{\bf p'}a^s_{\bf p}\rangle=\delta_{ss'}N_V({\bf p}, {\bf p'})+\mathcal{A}_{ss'}({\bf p}, {\bf p'}),
\end{eqnarray}
where $\mathcal{A}_{ss'}\neq 0$ when $s\neq s'$, which characterize certain projection in the spin space. When taking the spin sum, we assign
\begin{eqnarray}\nonumber
	&&\sum_s\xi_{s}\xi^{\dagger}_s=n\cdot\sigma=n\cdot\bar{\sigma}=I,
	\\ 
	&&\sum_{s,s'}\xi_{s}\mathcal{A}_{ss'}\xi^{\dagger}_{s'}=S({\bf p}, {\bf p'})\cdot\sigma,
\end{eqnarray}
where $S\cdot n=0$. Consequently, we find
\begin{widetext}
\begin{eqnarray}\label{vector_mtrx}
	\sum_su^s\left(q+\frac{p_-}{2}\right)\bar{u}^{s}\left(q-\frac{p_-}{2}\right)=
	\begin{pmatrix}
		\sqrt{\sigma\cdot\left(q+\frac{p_-}{2}\right)\bar{\sigma}\cdot\left(q-\frac{p_-}{2}\right)} &&
		\sqrt{\sigma\cdot\left(q+\frac{p_-}{2}\right)\sigma\cdot\left(q-\frac{p_-}{2}\right) } 
		\\
		\sqrt{\bar{\sigma}\cdot\left(q+\frac{p_-}{2}\right)\bar{\sigma}\cdot\left(q-\frac{p_-}{2}\right) }  &&
		\sqrt{\bar{\sigma}\cdot\left(q+\frac{p_-}{2}\right)\sigma\cdot\left(q-\frac{p_-}{2}\right) } 
	\end{pmatrix},
\end{eqnarray}
and
\begin{eqnarray}\label{axial_mtrx}
	\sum_{s,s'}u^s\left(q+\frac{p_-}{2}\right)\mathcal{A}_{ss'}\bar{u}^{s'}\left(q-\frac{p_-}{2}\right)=
	\begin{pmatrix}
		\sqrt{\sigma\cdot\left(q+\frac{p_-}{2}\right)}\sigma\cdot S\sqrt{\bar{\sigma}\cdot\left(q-\frac{p_-}{2}\right)} &&
		\sqrt{\sigma\cdot\left(q+\frac{p_-}{2}\right)}\sigma\cdot S\sqrt{\sigma\cdot\left(q-\frac{p_-}{2}\right) } 
		\\
		\sqrt{\bar{\sigma}\cdot\left(q+\frac{p_-}{2}\right)}\sigma\cdot S\sqrt{\bar{\sigma}\cdot\left(q-\frac{p_-}{2}\right) }  &&
		\sqrt{\bar{\sigma}\cdot\left(q+\frac{p_-}{2}\right)}\sigma\cdot S\sqrt{\sigma\cdot\left(q-\frac{p_-}{2}\right) } 
	\end{pmatrix}.
\end{eqnarray}
\end{widetext}

To compute the matrix elements above, we will employ the following tricks for Pauli matrices. We may write
\begin{eqnarray}
	q\cdot \sigma=m\exp\big(\hat{q}_{\perp}\cdot\sigma\theta\big),\quad \theta=\tanh^{-1}\left(\frac{|{\bf q}_{\perp}|}{E_q}\right),
\end{eqnarray}
where $\hat{q}^{\mu}_{\perp}=q^{\mu}_{\perp}/|{\bf q}_{\perp}|$, which yields
\begin{eqnarray}\nonumber
	\sqrt{q\cdot \sigma}&=&\sqrt{m}\Big(\cosh\frac{\theta}{2}+\hat{q}_{\perp}\cdot\sigma\sinh\frac{\theta}{2}\Big)
	\\
	&=&\sqrt{\frac{1}{2(E_q+m)}}\left((E_q+m)+q_{\perp}\cdot\sigma\right).
\end{eqnarray}
Hereafter we will use the subscripts $\perp$ to denote the components perpendicular to the frame vector $n^{\mu}$. That is, $V^{\mu}_{\perp}\equiv V^{\mu}-n\cdot Vn^{\mu}$ for arbitrary $V^{\mu}$.
We can now write
\begin{eqnarray}\nonumber
	&&\sqrt{\sigma\cdot q}=\sqrt{\frac{1}{2(E_q+m)}}(\chi_q+q_{\perp}\cdot\sigma),
	\\
	&&\sqrt{\bar{\sigma}\cdot q}=\sqrt{\frac{1}{2(E_q+m)}}(\chi_q-q_{\perp}\cdot\sigma),
\end{eqnarray}
where $\chi_q=E_q+m$. We then have to utilize the following parameterization,
\begin{eqnarray}\nonumber
	\tilde{v}^+_{\mu}\sigma^{\mu}
	&\equiv&(\chi_p+p_{\perp}\cdot\sigma)(\chi_{p'}+p'_{\perp}\cdot\sigma)
	\\\nonumber
	&=&\chi_p\chi_{p'}-p_{\perp}\cdot p'_{\perp}+(\chi_pp'_{\perp\mu}+\chi_{p'}p_{\perp\mu})\sigma^{\mu} 
	\\
	&&\quad-i\epsilon^{\mu\nu\alpha\beta}n_{\alpha}\sigma_{\beta}p_{\perp\mu}p'_{\perp\nu},
\end{eqnarray}
which gives
\begin{eqnarray}
	\tilde{v}^+\cdot n\approx 2E_q(E_q+m)+\mathcal{O}(p^2_-),
\end{eqnarray}
and
\begin{eqnarray}
	\tilde{v}^+_{\perp\mu}=2(E_q+m)q_{\perp\mu}-i\epsilon_{\mu\nu\alpha\beta}n^{\alpha}q^{\beta}p^{\nu}_{-}+\mathcal{O}(p^2_-).
\end{eqnarray}
Similarly, one finds
\begin{eqnarray}\nonumber
	\tilde{a}^+_{\mu}\sigma^{\mu}
	&\equiv&(\chi_p+p_{\perp}\cdot\sigma)S_{\perp}\cdot\sigma(\chi_{p'}+p'_{\perp}\cdot\sigma)
	\\\nonumber
	&=&\chi_p\chi_{p'}S_{\perp}\cdot\sigma-(\chi_pp'_{\perp\mu}+\chi_{p'}p_{\perp\mu}) S^{\mu}_{\perp}
	\\\nonumber
	&&-i\epsilon^{\mu\nu\alpha\beta}n_{\alpha}\sigma_{\beta}S_{\mu}(\chi_pp'_{\nu}-\chi_{p'}p_{\nu})+p_{\perp}\cdot p'_{\perp}S_{\perp}\cdot\sigma
	\\\nonumber
	&&-S_{\perp}\cdot p'_{\perp}p_{\perp}\cdot \sigma-S_{\perp}\cdot p_{\perp}p'_{\perp}\cdot \sigma
	-i\epsilon^{\mu\nu\alpha\beta}p_{\mu}p'_{\nu}n_{\alpha}S_{\beta}
	,\\
\end{eqnarray}
which gives
\begin{eqnarray}\nonumber
	\tilde{a}^+\cdot n\approx -2(E_q+m) q_{\perp}\cdot S_{\perp}
	+i\epsilon^{\mu\nu\alpha\beta}q_{\mu}p_{-\nu}n_{\alpha}S_{\beta}
	+\mathcal{O}(p^2_-),
	\\
\end{eqnarray}
and
\begin{eqnarray}\nonumber
	\tilde{a}^+_{\perp\mu}&=&2\big(m(E_q+m)S_{\perp\mu}-S_{\perp}\cdot q_{\perp}q_{\perp\mu}\big)
	\\\nonumber
&&-i\epsilon_{\mu\nu\alpha\beta}n^{\alpha}S^{\beta}\Big(\frac{q_{\perp}\cdot p_{-}}{E_q}q^{\nu}+(E_q+m)p_{-}^{\nu}\Big)+\mathcal{O}(p^2_-),
\\
\end{eqnarray}
where we use $p_-\cdot\partial_q\chi_q=-q_{\perp}\cdot p_{-}/E_q$.
On the other hand, 
we also introduce 
\begin{eqnarray}\nonumber
	&&\tilde{v}^-_{\mu}\sigma^{\mu}
	\equiv(\chi_p-p_{\perp}\cdot\sigma)(\chi_{p'}-p'_{\perp}\cdot\sigma),
	\\
	&&\tilde{a}^-_{\mu}\sigma^{\mu}
	\equiv(\chi_p-p_{\perp}\cdot\sigma)S_{\perp}\cdot\sigma(\chi_{p'}-p'_{\perp}\cdot\sigma).
\end{eqnarray}
In the end, up to $\mathcal{O}(p_-)$, we derive 
\begin{eqnarray}\label{av_tilde_mu_1}
	&&\tilde{v}^{\pm}\cdot n=(E_q+m)^2,
	\\\nonumber
	&&\tilde{a}^{\pm}\cdot n=\mp 2(E_q+m) q_{\perp}\cdot S_{\perp}
	+i\epsilon_{\mu\nu\alpha\beta}n^{\alpha}q^{\mu}p^{\nu}_{-}S_{\beta},
\end{eqnarray}
\begin{eqnarray}
	\tilde{v}^{\pm}_{\perp\mu}=\pm 2(E_q+m)q_{\perp\mu}- i\epsilon_{\mu\nu\alpha\beta}n^{\alpha}q^{\beta}p^{\nu}_{-},
\end{eqnarray}
and
\begin{eqnarray}\label{av_tilde_mu_3}
	\tilde{a}^{\pm}_{\perp\mu}&=&2\big(m(E_q+m)S_{\perp\mu}-S_{\perp}\cdot q_{\perp}q_{\perp\mu}\big)
	\\\nonumber
	&&\mp i\epsilon_{\mu\nu\alpha\beta}n^{\alpha}S^{\beta}\Big(\frac{q_{\perp}\cdot p_{-}}{E_q}q^{\nu}+(E_q+m)p_{-}^{\nu}\Big).
\end{eqnarray}

Let us focus on the off-diagonal terms in Eqs.~(\ref{vector_mtrx}) and (\ref{axial_mtrx}) associated with $\mathcal{V}_{\mu}\pm\mathcal{A}_{\mu}$. By utilizing Eqs.~(\ref{av_tilde_mu_1})-(\ref{av_tilde_mu_3}), it is found
\begin{eqnarray}\nonumber
&&\sigma\cdot(\mathcal{V}-\mathcal{A})
\\\nonumber
&&=\pi\int\frac{d^3p_-}{(2\pi)^3}\frac{e^{-ip_-\cdot X}\delta\left(q_0-E_q\right)}{2E_q(E_q+m)}
\left(\tilde{v}^+\cdot\sigma
N_V
+\tilde{a}^+\cdot\sigma
\right),
\\\nonumber
&&\bar{\sigma}\cdot(\mathcal{V}+\mathcal{A})
\\\nonumber
&&=\pi\int\frac{d^3p_-}{(2\pi)^3}\frac{e^{-ip_-\cdot X}\delta\left(q_0-E_q\right)}{2E_q(E_q+m)}
\left(\tilde{v}^-\cdot\sigma
N_V
+\tilde{a}^-\cdot\sigma
\right).\\
\end{eqnarray}
Recall that $N_V=N_V\left(\bm q+\frac{\bm p_-}{2},\bm q-\frac{\bm p_-}{2}\right)$ and $S_{\mu}=S_{\mu}\left(\bm q+\frac{\bm p_-}{2},\bm q-\frac{\bm p_-}{2}\right)$ in the integrands.
Thus, one obtains
\begin{eqnarray}\nonumber
	n\cdot\mathcal{A}&=&\frac{\pi}{2}\int\frac{d^3p_-}{(2\pi)^3}\frac{e^{-ip_-\cdot X}\delta\left(q_0-E_q\right)}{2E_q(E_q+m)}
	\\\nonumber
	&&\times n^{\mu}\left((\tilde{v}^-_{\mu}-\tilde{v}^+_{\mu})N_V
	+(\tilde{a}^-_{\mu}-\tilde{a}^+_{\mu})
	\right)
	\\
	&=&2\pi\delta(q^2-m^2)q_{\perp}\cdot \hat{S}_{\perp},
\end{eqnarray}
\begin{eqnarray}\nonumber
	\mathcal{A}_{\perp\mu}&=&-\frac{\pi}{2}\int\frac{d^3p_-}{(2\pi)^3}\frac{e^{-ip_-\cdot X}\delta\left(q_0-E_q\right)}{2E_q(E_q+m)}
	\\
	&&\times\left((\tilde{v}^-_{\perp\mu}+\tilde{v}^+_{\perp\mu})N_V
	+(\tilde{a}^-_{\perp\mu}+\tilde{a}^+_{\perp\mu})
	\right)
	\\\nonumber
	&=&2\pi\delta(q^2-m^2)\Big[\Big(\frac{\hat{S}_{\perp}\cdot q_{\perp}}{E_q+m}q_{\perp\mu}-m \hat{S}_{\perp\mu}\Big)
	\\
	&&-\epsilon_{\mu\nu\alpha\beta}\frac{n^{\alpha}q^{\beta}}{2(E_q+m)}\partial^{\nu}\tilde{f}_V\Big]
	,
\end{eqnarray}
\begin{eqnarray}\nonumber
	n\cdot\mathcal{V}&=&\frac{\pi}{2}\int\frac{d^3p_-}{(2\pi)^3}\frac{e^{-ip_-\cdot X}\delta\left(q_0-E_q\right)}{2E_q(E_q+m)}
	\\\nonumber
	&&\times n^{\mu}\left((\tilde{v}^-_{\mu}+\tilde{v}^+_{\mu})
	N_V
	+(\tilde{a}^-_{\mu}+\tilde{a}^+_{\mu})
	\right)
	\\\nonumber
	&=&2\pi\delta(q^2-m^2)\Big[E_q\tilde{f}_V+\frac{\epsilon^{\rho\nu\alpha\beta}}{2(E_q+m)}n_{\alpha}q_{\beta}\partial_{\nu}\hat{S}_{\rho}\Big]
	\\
	&=&2\pi\delta(q^2-m^2)E_qf_V
	,
\end{eqnarray}
and
\begin{eqnarray}\nonumber
	\mathcal{V}_{\perp\mu}&=&-\frac{\pi}{2}\int\frac{d^3p_-}{(2\pi)^3}\frac{e^{-ip_-\cdot X}\delta\left(q_0-E_q\right)}{2E_q(E_q+m)}
	\\\nonumber
	&&\times\left((\tilde{v}^-_{\perp\mu}-\tilde{v}^+_{\perp\mu})
	N_V
	+(\tilde{a}^-_{\perp\mu}-\tilde{a}^+_{\perp\mu})
	\right)
	\\\nonumber
	&=&2\pi\delta(q^2-m^2)\Big[\Big(q_{\perp\mu}\tilde{f}_V
	+\frac{\epsilon_{\mu\nu\alpha\beta}n^{\alpha}}{2(E_q+m)}
	\\\nonumber
	&&\times\Big(q^{\nu}\frac{q_{\perp}\cdot\partial}{E_q}+(E_q+m)\partial^{\nu}\Big)\hat{S}^{\beta}\Big]
	\\\nonumber
	&=&2\pi\delta(q^2-m^2)\Big[\Big(q_{\perp\mu}f_V(q,X)
	\\
	&&-\frac{\epsilon_{\mu\nu\alpha\beta}n^{\alpha}}{2E_q}\partial^{\nu}\Big(\frac{q\cdot \hat{S}_q^{\beta}}{(E_q+m)}-m\hat{S}^{\beta}\Big)\Big]
	,
\end{eqnarray}
where
\begin{eqnarray}\nonumber
	\tilde{f}_{V}(q,X)&\equiv&  \int\frac{d^3p_-}{(2\pi)^3}N_{V}\left(\bm q+\frac{\bm p_-}{2},\bm q-\frac{\bm p_-}{2}\right) e^{-ip_-\cdot X},
	\\\nonumber
	\hat{S}_{\mu}(q,X)&\equiv&  \int\frac{d^3p_-}{(2\pi)^3}S_{\mu}\left(\bm q+\frac{\bm p_-}{2},\bm q-\frac{\bm p_-}{2}\right) e^{-ip_-\cdot X}
	\\
\end{eqnarray}
with
\begin{eqnarray}\nonumber
	f_V(q,X)=\tilde{f}_V(q,X)+\frac{\epsilon^{\rho\nu\alpha\beta}}{2E_q(E_q+m)}n_{\alpha}q_{\beta}\partial_{\nu}
	\hat{S}_{\rho}(q,X).
	\\
\end{eqnarray}
Recall that $\hat{S}_{\mu}=\hat{S}_{\perp\mu}$.
In the computations above, we have employed the Schouten identity~\eqref{Schouten}.

Finally, by taking 
\begin{eqnarray}\label{Arel_a_and_S} \nonumber
	\hat{S}\cdot q_{\perp}=a\cdot nf_A,\quad \frac{q\cdot\hat{S}}{E_q+m}q_{\perp\mu}-m \hat{S}_{\mu}=a_{\perp\mu}f_A
	\\
\end{eqnarray}
and retrieving the $\hbar$ parameters, we obtain
\begin{eqnarray}\nonumber\label{free_case}
	\mathcal{A}_{\mu}&=&2\pi\delta(q^2-m^2)\Bigg(a_{\mu}f_A+\hbar\frac{\epsilon_{\mu\nu\alpha\beta}q^{\alpha}n^{\beta}}{2(q\cdot n+m)}\partial^{\nu}f_V\Bigg),
	\\\nonumber
	\mathcal{V}_{\mu}&=&2\pi\delta(q^2-m^2)\Bigg(q_{\mu}f_V+\hbar\frac{\epsilon_{\mu\nu\alpha\beta}n^{\beta}}{2(q\cdot n)}\partial^{\nu}(a^{\alpha}f_A)\Bigg),
	\\
\end{eqnarray}
where we replace $E_q$ by $q\cdot n$. Note that here $q\cdot a=q^2-m^2$ is indeed satisfied by taking $a\cdot n=q\cdot n+m$. One can now also decompose the spin four vector into $a^{\mu}f_A=(q^{\mu}+mn^{\mu})f_A-\hat{S}^{\mu}$. 

In the presence of arbitrary background fields, the analytic solution for Dirac wave functions is unknown. Consequently, we generalize the free solution for axial WFs based on the solution in Eq.~(\ref{free_case}) and its connection to the massless result for Weyl fermions. We hence conclude
\begin{eqnarray}\label{Axial_sol}\nonumber
	\mathcal{A}_{\mu}&=&2\pi\delta(q^2-m^2) \Big(a_{\mu}f_A+\hbar S^{\mu\nu}_{m(n)}\Delta^{\nu}f_V\Big)
	\\
	&&+\hbar\tilde{F}_{\mu\nu}q^{\nu}2\pi\delta'(q^2-m^2)f_V,
\end{eqnarray} 
by replacing the $\partial_{\nu}$ operator with the $\Delta^{\nu}$ operator in the magnetization-current term,
where
\begin{eqnarray}\label{Aspin_tensor_m}
	S^{\mu\nu}_{m(n)}=\frac{\epsilon^{\mu\nu\alpha\beta}q_{\alpha}n_{\beta}}{2(q\cdot n+m)}
	. 
\end{eqnarray} 

\section{Scalar/Axial-Vector Kinetic Equations}\label{sec:Scalar/Axial-VectorKineticEquations}
Given the perturbative solution for $\mathcal{V}^{\mu}$ up to $\mathcal{O}(\hbar)$ in Eq.~(\ref{AY_sol_V1}), we first derive the scalar kinetic equation (SKE) from $\Delta\cdot\mathcal{V}=0$ in Eq.~(\ref{Aeqv1}). In the derivation, we assume that the frame vector $n_{\mu}$ is independent of the momentum $q$. 
By performing straightforward computations, we find
\begin{widetext}
\begin{eqnarray}\nonumber\label{div_vector1}
	\Delta\cdot \mathcal{V}_1&=&\frac{2\pi\delta(q^2-m^2)\epsilon^{\mu\nu\rho\sigma}}{2n\cdot q}\Bigg(\Bigg(\frac{E_{\mu}n_{\nu}a_{\sigma}}{q\cdot n}+(q\cdot n)\partial_{\mu}\left(\frac{n_{\nu}a_{\sigma}}{q\cdot n}\right)-n_{\nu}(\partial_{\mu}a_{\sigma})\Bigg)\Delta_{\rho}
	+n_{\nu}a_{\sigma}(\partial_{\mu}F_{\beta\rho})\partial_q^{\beta}
	\Bigg)f_A
	\\\nonumber
	&&+\frac{2\pi\delta(q^2-m^2)\epsilon^{\mu\nu\rho\sigma}f_{A}}{2n\cdot q}\Bigg(\left((\partial_{\mu}n_{\nu})+\frac{E_{\mu}-(\partial_{\mu}q\cdot n)}{q\cdot n}n_{\nu}\right)\Delta_{\rho}+n_{\nu}(\partial_{\mu}F_{\beta\rho})\partial_q^{\beta}\Bigg)a_{\sigma}
	\\\nonumber
	&&-\frac{2\pi\delta'(q^2-m^2)}{n\cdot q}B^{\mu}\big(q\cdot\Delta \tilde{a}_{\mu}+F_{\nu\mu}\tilde{a}^{\nu}\big)
	+\frac{2\pi\tilde{F}^{\mu\nu}\delta(q^2-m^2)f_A}{q\cdot n}\left((\partial_{\mu}n_{\nu})+\frac{E_{\mu}-(\partial_{\mu}q\cdot n)}{q\cdot n}n_{\nu}\right)
	\\\nonumber
	&=&2\pi\delta(q^2-m^2)\Bigg(\frac{E_{\mu}S^{\mu\nu}_{a(n)}}{q\cdot n}\Delta_{\nu} + S^{\mu\nu}_{a(n)}(\partial_{\mu}F_{\rho\nu})\partial^{\rho}_q
	+\big(\partial_{\mu}S^{\mu\nu}_{a(n)}\big)\Delta_{\nu}\Bigg)f_A
	-\frac{2\pi\delta'(q^2-m^2)}{n\cdot q}B^{\mu}\big(q\cdot\Delta \tilde{a}_{\mu}+F_{\nu\mu}\tilde{a}^{\nu}\big)
	\\
	&&+\pi\delta(q^2-m^2)\epsilon^{\mu\nu\alpha\beta}\Bigg(\frac{n_{\beta}}{q\cdot n}\Big((\partial_{\nu}a_{\alpha})\Delta_{\mu}+(\partial_{\mu}F_{\rho\nu})(\partial_q^{\rho}a_{\alpha})\Big)
	+\Delta_{\mu}\left(\frac{n^{\beta}}{q\cdot n}\right)((\Delta_{\nu}a_{\alpha})+F_{\nu\alpha})
	\Bigg)f_A,
\end{eqnarray}
where the electric/magnetic fields are defined in terms of $n^{\mu}$,
\begin{eqnarray}
	&&\tilde{F}_{\mu\nu}=\frac{1}{2}\epsilon_{\mu\nu\alpha\beta}F^{\alpha\beta}=\epsilon_{\mu\nu\alpha\beta}E^{\alpha}n^{\beta}+B_{\mu}n_{\nu}-B_{\nu}n_{\mu}, 
	\quad
F_{\mu\nu}=-\epsilon_{\mu\nu\alpha\beta}B^{\alpha}n^{\beta}+E_{\mu}n_{\nu}-E_{\nu}n_{\mu}.
\end{eqnarray}
From Eq.~(\ref{div_vector1}), we derive the SKE
\begin{eqnarray}\nonumber\label{vector_CKT}
	0&=&\delta(q^2-m^2)\Bigg[q\cdot\Delta f_V+\frac{\hbar}{2}\Bigg(\frac{E_{\mu}S^{\mu\nu}_{a(n)}}{q\cdot n}\Delta_{\nu} + S^{\mu\nu}_{a(n)}(\partial_{\mu}F_{\rho\nu})\partial^{\rho}_q
	+\big(\partial_{\mu}S^{\mu\nu}_{a(n)}\big)\Delta_{\nu}\Bigg)f_A\Bigg]
	\\\nonumber
	&&+\frac{\hbar\delta(q^2-m^2)\epsilon^{\mu\nu\alpha\beta}}{4}\Bigg(\frac{n_{\beta}}{q\cdot n}\Big((\partial_{\nu}a_{\alpha})\Delta_{\mu}+(\partial_{\mu}F_{\rho\nu})(\partial_q^{\rho}a_{\alpha})\Big)
	+\Delta_{\mu}\left(\frac{n^{\beta}}{q\cdot n}\right)((\Delta_{\nu}a_{\alpha})+F_{\nu\alpha})
	\Bigg)f_A
	\\
	&&-\frac{\hbar\delta'(q^2-m^2)}{2q\cdot n}B^{\mu}\big(q\cdot\Delta (a_{\mu}f_A)+F_{\nu\mu}a^{\nu}f_A\big),
\end{eqnarray}
\end{widetext}
where
\begin{eqnarray}
	S^{\mu\nu}_{a(n)}=\frac{\epsilon^{\mu\nu\alpha\beta}a_{\alpha}n_{\beta}}{2q\cdot n}.
\end{eqnarray}

Next, we may derive the axial-vector equation (AKE) from the perturbative solution of $\mathcal{A}_{\mu}$ up to $\mathcal{O}(\hbar)$ in Eqs.~(\ref{Axial_sol}) and (\ref{AeqA3}) in the master equations. The computations will be more complicated than the case for SKE but straightforward. Nevertheless, in order to make a direct comparison with the massless CKT, the underlying strategy is to isolate the $\hbar$ terms proportional to $q^{\mu}$ and the other terms explicitly proportional to $m$ since we expect that the AKE should reduce to $q^{\mu}$ multiplied by CKT in the massless limit as foreseen from the off-shell BMT equation. 

From Eqs.~(\ref{Axial_sol}) and (\ref{AeqA3}), we obtain
\begin{widetext}
\begin{eqnarray}\nonumber\label{CKT_axial_meq}
	0&=&\delta(q^2-m^2)\Big[
	\Big(a^{\mu}q\cdot\Delta f_A+f_A\big(q\cdot\Delta a^{\mu}+F^{\nu\mu}a_{\nu}\big)\Big)
	+\hbar\big(q\cdot\Delta (S^{\mu\nu}_{m(n)}\Delta_{\nu}f_V)+F^{\nu\mu}S_{m(n)\nu\rho}\Delta^{\rho}f_V\big)
	\\
	&&
	-\hbar\epsilon^{\mu\nu\rho\sigma}(\partial_{\sigma}F_{\beta\nu})q_{\rho}\partial_q^{\beta}f_V
	\Big]
	+\hbar\delta'(q^2-m^2)\tilde{F}^{\mu\nu}q_{\nu}q\cdot\Delta f_V.
\end{eqnarray}
\end{widetext}
We then rearrange this equation in light of the aforementioned strategy to obtain the form for comparison with the CKT when $m=0$.
We shall first evaluate
\begin{eqnarray}\nonumber\label{DS_m_full}
	&&\delta(q^2-m^2)q\cdot\Delta\big(S^{\mu\nu}_{m(n)}\Delta_{\nu}f_V\big)
	\\\nonumber
	&&=\delta(q^2-m^2)\big((q\cdot\Delta S^{\mu\nu}_{m(n)})\Delta_{\nu}f_V+ S^{\mu\nu}_{m(n)}q\cdot\Delta (\Delta_{\nu}f_V)\big)
	\\\nonumber
	&&=\Big[\delta(q^2-m^2)\Big(\big(q\cdot\Delta S^{\mu\nu}_{m(n)}\big)\Delta_{\nu}-S^{\mu\nu}_{m(n)}F_{\rho\nu}\Delta^{\rho}
	\\\nonumber
	&&\quad
	+S^{\mu\nu}_{m(n)}\big((q\cdot\partial F_{\beta\nu})-q^{\rho}(\partial_{\nu}F_{\beta\rho})\big)\partial_q^{\beta}
	\Big)
	\\
	&&\quad
	-2\delta'(q^2-m^2)q^{\rho}F_{\rho\nu}S^{\mu\nu}_{m(n)}q\cdot\Delta\Big] f_V,
\end{eqnarray}
where we take
\begin{eqnarray}\nonumber
	&&\delta(q^2-m^2)q\cdot\Delta(\Delta_{\nu}f_V)
	\\\nonumber
	&&=\delta(q^2-m^2)\Big((q\cdot\Delta\Delta_{\nu})f_V+\Delta_{\nu}(q\cdot\Delta f_V)-(\Delta_{\nu}q\cdot\Delta)f_V\Big)
	\\\nonumber
	&&=\delta(q^2-m^2)\Big((q\cdot\Delta\Delta_{\nu})f_V-(\Delta_{\nu}q\cdot\Delta)f_V\Big)
	\\\nonumber
	&&\quad+\Delta_{\nu}(\delta(q^2-m^2)q\cdot\Delta f_V)
	-(\Delta_{\nu}\delta(q^2-m^2))(q\cdot\Delta f_V).
	\\
\end{eqnarray}
For the first component in Eq.~(\ref{DS_m_full}), we find
\begin{widetext}
\begin{eqnarray}\nonumber
	&&\delta(q^2-m^2)\big(q\cdot\Delta S^{\mu\nu}_{m(n)}\big)\Delta_{\nu}f_V
	\\\nonumber
	&&
	=\delta(q^2-m^2)\Bigg[q^{\mu}(\Delta_{\alpha}S_{m(n)}^{\alpha\nu})+(\Delta_{\alpha}S^{\mu\alpha}_{m(n)})q^{\nu}
	+\epsilon^{\mu\nu\alpha\sigma}\Bigg(\frac{q^{\rho}F_{\rho\alpha}n_{\sigma}}{2(q\cdot n+m)}+q^2\Bigg(\partial_{\alpha}\frac{n_{\sigma}}{2(q\cdot n+m)}\Bigg)\Bigg)
	+\frac{\epsilon^{\mu\nu\alpha\sigma}q^2n_{\sigma}E_{\alpha}}{2(q\cdot n+m)^2}
	\\\nonumber
	&&\quad +\epsilon^{\mu\nu\rho\alpha}\Bigg(\frac{q\cdot nF_{\rho\alpha}}{2(q\cdot n+m)}+q_{\rho}\Bigg(\partial_{\alpha}\frac{q\cdot n}{2(q\cdot n+m)}\Bigg)\Bigg)
	+\frac{\epsilon^{\mu\nu\rho\alpha}q\cdot nq_{\rho}E_{\alpha}}{2(q\cdot n+m)^2}
	\Bigg]\Delta_{\nu}f_V
	\\\nonumber
	&&=\delta(q^2-m^2)\Bigg[q^{\mu}(\partial_{\alpha}S_{m(n)}^{\alpha\nu})+q^{\mu}\frac{S^{\alpha\nu}_{m(n)}E_{\alpha}}{q\cdot n+m}
	+\frac{\big(q\cdot n\tilde{F}^{\mu\nu}+q^{\mu}B^{\nu}\big)}{2(q\cdot n+m)}
	+\frac{\epsilon^{\mu\nu\rho\alpha}q_{\rho}E_{\alpha}}{2(q\cdot n+m)}
	+\frac{\epsilon^{\mu\nu\alpha\beta}m^2}{2(q\cdot n+m)}(\partial_{\alpha}n_{\beta})
	\\
	&&\quad+\frac{\epsilon^{\mu\nu\alpha\beta}(m^2n_{\beta}+mq_{\beta})}{2(q\cdot n+m)^2}\big(E_{\alpha}-\partial_{\alpha}(q\cdot n)\big)\Bigg]\Delta_{\nu}f_V+\mathcal{O}(\hbar),
\end{eqnarray}
\end{widetext}
where we use
\begin{eqnarray}\nonumber
	&&\epsilon^{\mu\nu\alpha\sigma}F_{\rho\alpha}=\delta^{\mu}_{\rho}B^{[\sigma}n^{\nu]}+\delta^{\sigma}_{\rho}B^{[\nu}n^{\mu]}+\delta^{\nu}_{\rho}B^{[\mu}n^{\sigma]}+\epsilon^{\mu\nu\alpha\sigma}E_{[\rho}n_{\alpha]}
	\\
	&&=\delta^{\mu}_{\rho}\tilde{F}^{\sigma\nu}+\delta^{\sigma}_{\rho}\tilde{F}^{\nu\mu}+\delta^{\nu}_{\rho}\tilde{F}^{\mu\sigma},
\end{eqnarray}
and
\begin{eqnarray}\nonumber
	\Delta_{\alpha}S_{m(n)}^{\alpha\nu}&=&\partial_{\alpha}S_{m(n)}^{\alpha\nu}
	+\epsilon^{\alpha\nu\rho\sigma}F_{\beta\alpha}\partial_q^{\beta}\Big(\frac{q_{\rho}n_{\sigma}}{2(q \cdot n+m)}\Big)
	\\
	&=&\partial_{\alpha}S_{m(n)}^{\alpha\nu}+\frac{B^{\nu}}{q\cdot n+m}+\frac{S^{\alpha\nu}_{m(n)}E_{\alpha}}{q\cdot n+m}.
\end{eqnarray}
Subsequently, the second component in Eq.~(\ref{DS_m_full}) reads
\begin{eqnarray}\nonumber
	&&-\delta(q^2-m^2)S^{\mu\nu}_{m(n)}F_{\rho\nu}\Delta^{\rho}f_V
	\\\nonumber
	&&=-\frac{\delta(q^2-m^2)}{2(q\cdot n+m)}\epsilon^{\mu\nu\alpha\beta}q_{\alpha}n_{\beta}F_{\rho\nu}\Delta^{\rho}f_V
	\\\nonumber
	&&=\frac{\delta(q^2-m^2)}{2(q\cdot n+m)}\big(\delta^{\mu}_{\rho}\tilde{F}^{\beta\alpha}+\delta^{\alpha}_{\rho}\tilde{F}^{\mu\beta}+\delta^{\beta}_{\rho}\tilde{F}^{\alpha\mu}\big)q_{\alpha}n_{\beta}\Delta^{\rho}f_V
	\\
	&&=\frac{\delta(q^2-m^2)}{2(q\cdot n+m)}\big(-q\cdot B\Delta^{\mu}+q_{\alpha}\tilde{F}^{\alpha\mu}n\cdot\Delta\big)f_V.
\end{eqnarray}
Next, the third component in Eq.~(\ref{DS_m_full}) can be written as
\begin{eqnarray}\nonumber
	&&\delta(q^2-m^2)S^{\mu\nu}_{m(n)}\big((q\cdot\partial F_{\beta\nu})-q^{\rho}(\partial_{\nu}F_{\beta\rho})\big)\partial_q^{\beta}f_V
	\\\nonumber
	&&=\delta(q^2-m^2)S^{\mu\nu}_{m(n)}(q\cdot\partial F_{\beta\nu})\partial_q^{\beta}f_V
	-\frac{\delta(q^2-m^2)}{2(q\cdot n+m)}\big(q^{\mu}\epsilon^{\rho\nu\alpha\sigma}
	\\\nonumber
	&&\quad+q^{\nu}\epsilon^{\mu\rho\alpha\sigma} +q^{\alpha}\epsilon^{\mu\nu\rho\sigma}+q^{\sigma}\epsilon^{\mu\nu\alpha\rho}\big)q_{\alpha}n_{\sigma}(\partial_{\nu}F_{\beta\rho})\partial_q^{\beta}f_V
	\\\nonumber
	&&=
	-\delta(q^2-m^2)\Big(q^{\mu}S^{\rho\nu}_{m(n)}(\partial_{\nu}F_{\beta\rho})+\frac{m^2\epsilon^{\mu\nu\rho\sigma}n_{\sigma}}{2(q\cdot n+m)}(\partial_{\nu}F_{\beta\rho})
	\\
	&&\quad-\Big(1-\frac{m}{q\cdot n+m}\Big) \frac{\epsilon^{\mu\nu\rho\sigma}q_{\rho}}{2}(\partial_{\sigma}F_{\beta\nu})\Big)\partial_q^{\beta}f_V.
\end{eqnarray}
For the forth component in Eq.~(\ref{DS_m_full}), it is found
\begin{eqnarray}\nonumber
	&&-\delta'(q^2-m^2)q^{\rho}F_{\rho\nu}S^{\mu\nu}_{m(n)}q\cdot\Delta f_V
	\\\nonumber
	&&=-\frac{\delta'(q^2-m^2)}{2(q\cdot n+m)}q^{\rho}F_{\rho\nu}\epsilon^{\mu\nu\alpha\beta}q_{\alpha}n_{\beta}q\cdot\Delta f_V
	\\\nonumber
	&&=-\frac{\delta'(q^2-m^2)}{2(q\cdot n+m)}F_{\rho\nu}(q^{\mu}\epsilon^{\rho\nu\alpha\beta}+q^{\nu}\epsilon^{\mu\rho\alpha\beta}+q^{\alpha}\epsilon^{\mu\nu\rho\beta}
	\\\nonumber
	&&\quad+q^{\beta}\epsilon^{\mu\nu\alpha\rho})q_{\alpha}n_{\beta}q\cdot\Delta f_V
	\\\nonumber
	&&=-\Bigg[q^{\mu}q\cdot B\frac{\delta'(q^2-m^2)}{q\cdot n+m} 
	-\delta'(q^2-m^2)q^{\rho}F_{\rho\nu}S^{\mu\nu}_{m(n)}
	\\
	&&-\frac{\delta'(q^2-m^2)}{q\cdot n+m}\tilde{F}^{\mu\beta}(q^2n_{\beta}-q\cdot nq_{\beta})\Bigg]q\cdot\Delta f_V,
\end{eqnarray}
which yields
\begin{eqnarray}\nonumber
	&&-\delta'(q^2-m^2)q^{\rho}F_{\rho\nu}S^{\mu\nu}_{m(n)}\frac{q\cdot\Delta f_V}{2}
	\\\nonumber
	&&=-\Bigg[q^{\mu}q\cdot B\frac{\delta'(q^2-m^2)}{q\cdot n+m} 
	-\frac{\delta'(q^2-m^2)}{q\cdot n+m}(m^2B^{\mu}
	\\\nonumber
	&&\quad-q\cdot nq_{\beta}\tilde{F}^{\mu\beta})\Bigg]q\cdot\Delta f_V
	+\mathcal{O}(\hbar)
	\\\nonumber
	&&=-\frac{\delta'(q^2-m^2)}{2}\Bigg[\frac{q^{\mu}q\cdot B}{q\cdot n+m} 
	-\frac{m(mB^{\mu}+q_{\beta}\tilde{F}^{\mu\beta})}{q\cdot n+m}
	\\
	&&\quad+q_{\nu}\tilde{F}^{\mu\nu}\Bigg]q\cdot\Delta f_V
	+\mathcal{O}(\hbar)
	.
\end{eqnarray}

On the other hand, one finds
\begin{eqnarray}
	&&\delta(q^2-m^2)F^{\nu\mu}S_{m(n)\nu\rho}\Delta^{\rho}f_V
	\\\nonumber
	&&=\delta(q^2-m^2)\frac{\epsilon_{\rho\nu\alpha\sigma}F^{\mu\alpha}}{2(q\cdot n+m)}q^{\sigma}n^{\nu}\Delta^{\rho}f_V
	\\\nonumber
	&&=\frac{\delta(q^2-m^2)}{2(q\cdot n+m)}\Big[q\cdot B\Delta^{\mu}-q^{\mu}B\cdot\Delta+n^{\mu}q^{\rho}\tilde{F}_{\nu\rho}\Delta^{\nu}
	\Big]f_V.
\end{eqnarray}
Combining all pieces together, we acquire
\begin{widetext}
\begin{align}\nonumber
	&\delta(q^2-m^2)\Big[q\cdot\Delta \big(S^{\mu\nu}_{m(n)}\Delta_{\nu}f_V\big)+F^{\nu\mu}S_{m(n)\nu\rho}\Delta^{\rho}f_V\Big]
	+\delta'(q^2-m^2)\tilde{F}^{\mu\nu}q_{\nu}q\cdot\Delta f_V
	-\frac{\epsilon^{\mu\nu\rho\sigma}}{2}(\partial_{\sigma}F_{\beta\nu})q_{\rho}\partial_q^{\beta}f_V
	\\\nonumber
	&=\delta(q^2-m^2)\Bigg[q^{\mu}(\partial_{\alpha}S_{m(n)}^{\alpha\nu})+q^{\mu}\frac{S^{\alpha\nu}_{m(n)}E_{\alpha}}{q\cdot n+m}
	+\frac{\big(q\cdot n\tilde{F}^{\mu\nu}+q_{\rho}\tilde{F}^{\rho\mu}n^{\nu}+n^{\mu}q_{\rho}\tilde{F}^{\nu\rho}\big)}{2(q\cdot n+m)}
	+\frac{\epsilon^{\mu\nu\rho\alpha}q_{\rho}E_{\alpha}}{2(q\cdot n+m)}
	\\\nonumber
	&\quad+\frac{\epsilon^{\mu\nu\alpha\beta}(m^2n_{\beta}+mq_{\beta})}{2(q\cdot n+m)^2}\big(E_{\alpha}-\partial_{\alpha}(q\cdot n)\big)\Bigg]\Delta_{\nu}f_V
	-\delta'(q^2-m^2)\Bigg[\frac{q^{\mu}q\cdot B}{q\cdot n+m} 
	-\frac{m(mB^{\mu}+q_{\beta}\tilde{F}^{\mu\beta})}{q\cdot n+m}\Bigg]q\cdot\Delta f_V
	\\\nonumber
	&\quad
	+\delta(q^2-m^2)\Bigg[q^{\mu}S^{\rho\nu}_{m(n)}(\partial_{\rho}F_{\beta\nu})+\frac{\epsilon^{\mu\nu\rho\sigma}m(mn_{\rho}+q_{\rho})}{2(q\cdot n+m)}(\partial_{\nu}F_{\beta\sigma})\Bigg]\partial_q^{\beta}f_V
	\\\nonumber
	&=q^{\mu}\Bigg\{\delta(q^2-m^2)\Bigg[(\partial_{\alpha}S_{m(n)}^{\alpha\nu})\Delta_{\nu}+\frac{S^{\alpha\nu}_{m(n)}E_{\alpha}\Delta_{\nu}}{q\cdot n+m}
	+S^{\rho\nu}_{m(n)}(\partial_{\rho}F_{\beta\nu})\partial_q^{\beta}\Bigg]
	-\delta'(q^2-m^2)\frac{q\cdot B}{q\cdot n+m}q\cdot\Delta\Bigg\}f_V
	\\\nonumber
	&\quad
	+m\Bigg\{\delta(q^2-m^2)\frac{\epsilon^{\mu\nu\alpha\beta}(mn_{\beta}+q_{\beta})}{2(q\cdot n+m)}\Bigg(\frac{\big(E_{\alpha}-\partial_{\alpha}(q\cdot n)\big)}{q\cdot n+m}\Delta_{\nu}-(\partial_{\nu}F_{\rho\alpha})\partial_q^{\rho}\Bigg)
	\\
	&\quad+\delta'(q^2-m^2)\frac{(mB^{\mu}+q_{\beta}\tilde{F}^{\mu\beta})}{q\cdot n+m}q\cdot\Delta\Bigg\} f_V
	.
\end{align}
\end{widetext}
To obtain the last equality above, we apply
\begin{eqnarray}\nonumber
	&&q\cdot n\tilde{F}^{\mu\nu}+q_{\rho}\tilde{F}^{\rho\mu}n^{\nu}+n^{\mu}q_{\rho}\tilde{F}^{\nu\rho}+\epsilon^{\mu\nu\rho\alpha}q_{\rho}E_{\alpha}
	\\\nonumber
	&&=q\cdot n(B^{[\mu}n^{\nu]}+\epsilon^{\mu\nu\alpha\beta}E_{\alpha}n_{\beta})
	+q\cdot Bn^{\mu}n^{\nu}-q\cdot nB^{\mu}n^{\nu}
	\\\nonumber
	&&\quad+\epsilon^{\rho\mu\alpha\beta}q_{\rho}E_{\alpha}n_{\beta}n^{\nu}
	+n^{\mu}(B^{\nu}q\cdot n-q\cdot Bn^{\nu})
	\\\nonumber
	&&\quad+n^{\mu}\epsilon^{\nu\rho\alpha\beta}q_{\rho}E_{\alpha}n_{\beta}+\epsilon^{\mu\nu\rho\alpha}q_{\rho}E_{\alpha}
	\\\nonumber
	&&=\epsilon^{\mu\nu\alpha\beta}(q\cdot nE_{\alpha}n_{\beta}-E_{\alpha}q_{\beta})+q_{\rho}E_{\alpha}n_{\beta}(\epsilon^{\rho\mu\alpha\beta}n^{\nu}
	-\epsilon^{\rho\nu\alpha\beta}n^{\mu})
	\\
	&&=0,
\end{eqnarray}
where we also use 
\begin{eqnarray}\nonumber
	q_{\rho}E_{\alpha}n_{\beta}\epsilon^{\rho\nu\alpha\beta}n^{\mu}&=&(q\cdot n)E_{\alpha}n_{\beta}\epsilon^{\mu\nu\alpha\beta}+q_{\rho}E_{\alpha}n_{\beta}\epsilon^{\rho\mu\alpha\beta}n^{\nu}
	\\
	&&+\epsilon^{\rho\nu\alpha\mu}q_{\rho}E_{\alpha}.
\end{eqnarray}

From Eq.~(\ref{CKT_axial_meq}), the AKE takes the form
\begin{widetext}
\begin{eqnarray}\nonumber\label{AKE_final}
	0&=&\delta(q^2-m^2)
	\Big(a^{\mu}q\cdot\Delta f_A+f_A\big(q\cdot\Delta a^{\mu}+F^{\nu\mu}a_{\nu}\big)\Big)
	+\hbar q^{\mu}\Bigg\{\delta(q^2-m^2)\Bigg[(\partial_{\alpha}S_{m(n)}^{\alpha\nu})\Delta_{\nu}+\frac{S^{\alpha\nu}_{m(n)}E_{\alpha}\Delta_{\nu}}{q\cdot n+m}
	\\\nonumber
	&&+S^{\rho\nu}_{m(n)}(\partial_{\rho}F_{\beta\nu})\partial_q^{\beta}\Bigg]
	-\delta'(q^2-m^2)\frac{q\cdot B}{q\cdot n+m}q\cdot\Delta\Bigg\}f_V
	\\\nonumber
	&&
	+\hbar m\Bigg\{\delta(q^2-m^2)\frac{\epsilon^{\mu\nu\alpha\beta}(mn_{\beta}+q_{\beta})}{2(q\cdot n+m)}\Bigg(\frac{\big(E_{\alpha}-\partial_{\alpha}(q\cdot n)\big)}{q\cdot n+m}\Delta_{\nu}-(\partial_{\nu}F_{\rho\alpha})\partial_q^{\rho}\Bigg)
	\\
	&&+\delta'(q^2-m^2)\frac{(mB^{\mu}+q_{\beta}\tilde{F}^{\mu\beta})}{q\cdot n+m}q\cdot\Delta\Bigg\} f_V
	.
\end{eqnarray}
\end{widetext}
\section{Spin-Hall Effect}\label{sec:Spin-Hall Effect}
We show how Eq.~(\ref{AKE_final}) reveals a spin Hall effect in a non-relativistic case. Assuming $E^{\mu}$ and $n^{\mu}$ are constant and approximating $q^{\mu}\approx mn^{\nu}$, Eq.~(\ref{AKE_final}) reduces to
\begin{eqnarray}\label{non_rel_CKT_E}\nonumber
	&&\delta(q^2-m^2)\Big(\Box^{\mu\nu}\tilde{a}_{\nu}+\frac{\hbar}{4}\epsilon^{\mu\nu\alpha\beta}E_{\alpha}n_{\beta}\Delta_{\nu}f_V\Big)
	\\
	&&+\frac{\hbar}{2}\delta'(q^2-m^2)\epsilon^{\mu\nu\alpha\beta}E_{\alpha}n_{\beta}q_{\nu}q\cdot\Delta f_V\approx 0,
\end{eqnarray} 
where $\Box^{\mu\nu}=\eta^{\mu\nu}q\cdot\Delta+F^{\nu\mu}$ and $\tilde{a}^{\nu}=a^{\nu} f_A$. By using
\begin{eqnarray}
	&&\delta'(q^2-m^2)\epsilon^{\mu\nu\alpha\beta}E_{\alpha}n_{\beta}q_{\nu}q\cdot\Delta f_V
	\\\nonumber
	&&=-\frac{\delta(q^2-m^2)}{2}\epsilon^{\mu\nu\alpha\beta}E_{\alpha}n_{\beta}\big(\Delta_{\nu}+q\cdot\Delta\partial_{q\nu}\big)f_V+\mathcal{O}(\hbar),
\end{eqnarray}
Eq.~(\ref{non_rel_CKT_E}) becomes
\begin{eqnarray}\nonumber
	\delta(q^2-m^2)\Big[q\cdot\Delta\Big(\tilde{a}^{\mu}-\frac{\hbar\epsilon^{\mu\nu\alpha\beta}}{4}E_{\alpha}n_{\beta}\partial_{q\nu}f_V\Big)+F^{\nu\mu}\tilde{a}_{\nu}\Big]\approx 0,
	\\
\end{eqnarray}
which can be written as
\begin{eqnarray}\label{approx_CKT_E}
	&&n\cdot\Delta\Big(\tilde{a}^{\mu}-\frac{\hbar\epsilon^{\mu\nu\alpha\beta}}{4}E_{\alpha}n_{\beta}\partial_{q\nu}f_V\Big)
	\\\nonumber
	&&=(n\cdot\partial+E_{\rho}\partial_q^{\rho})\Big(\tilde{a}^{\mu}-\frac{\hbar\epsilon^{\mu\nu\alpha\beta}}{4}E_{\alpha}n_{\beta}\partial_{q\nu}f_V\Big)
	\approx 0,
\end{eqnarray}
by further dropping the $\mathcal{O}(1/m)$ suppression terms. On the other hand, in such a limit, the axial WF approximately reads
\begin{eqnarray}\nonumber
	\mathcal{A}_{\mu}&\approx& 2\pi\delta(q^2-m^2) a_{\mu}f_A
	\\
	&&+\hbar\pi\epsilon_{\mu\nu\alpha\beta}E^{\alpha}n^{\beta}(\partial^{\nu}_{q}\delta(q^2-m^2))f_V.
\end{eqnarray}
In a stationary state such that
\begin{eqnarray}
	\tilde{a}^{\mu}=\frac{\hbar}{4}\epsilon^{\mu\nu\alpha\beta}E_{\alpha}n_{\beta}\partial_{q\nu}f_V,
\end{eqnarray}
we find
\begin{eqnarray}
	J^{\mu}_{5}&=&4\int\frac{d^4q}{(2\pi)^4}\mathcal{A}^{\mu}\notag\\
	&\approx& 4\pi\int\frac{d^4q}{(2\pi)^4}\delta(q^2-m^2)
	\Big(2\tilde{a}^{\mu}-\hbar\epsilon^{\mu\nu\alpha\beta}E_{\alpha}n_{\beta}\partial_{q\nu}f_V\Big)\notag\\
	&=&-2\pi\hbar\epsilon^{\mu\nu\alpha\beta}E_{\alpha}n_{\beta}\int\frac{d^4q}{(2\pi)^4}\delta(q^2-m^2)\partial_{q\nu}f_V.
\end{eqnarray}

\section{Frame Independence}\label{sec:FrameIndependence}
In this section, we derive the modified frame transformation upon $f_V$ and $a^{\mu} f_A$ at $\mathcal{O}(\hbar)$ to ensure the frame independence of $\mathcal{V}^{\mu}$ and $\mathcal{A}^{\mu}$. Recall that the explicit form of $\mathcal{V}^{\mu}$ and $\mathcal{A}^{\mu}$ in an arbitrary frame $n^{\mu}$ reads
\begin{widetext}
\begin{eqnarray}
	\mathcal{V}^{\mu}&=&2\pi\delta(q^2-m^2)\Bigg[q^{\mu}f^{(n)}_V
	+\frac{\hbar\epsilon^{\mu\nu\rho\sigma}n_{\nu}}{2q\cdot n}
	\big(\Delta_{\rho}(a^{(n)}_{\sigma}f^{(n)}_A)+F_{\rho\sigma}f^{(n)}_A\big)
	\Bigg]+2\pi\hbar\tilde{F}^{\mu\nu}a^{(n)}_{\nu}\delta'(q^2-m^2)f^{(n)}_A,
	\\
	\mathcal{A}^{\mu}&=&2\pi\delta(q^2-m^2)\Bigg[a^{(n)\mu}f_A^{(n)}
	+\frac{\hbar\epsilon^{\mu\nu\alpha\beta}q_{\alpha}n_{\beta}}{2(q\cdot n+m)}\Delta_{\nu}f_V^{(n)}
	\Bigg]+2\pi\hbar\tilde{F}^{\mu\nu}q_{\nu}\delta'(q^2-m^2)f^{(n)}_V,
\end{eqnarray}  
\end{widetext}
where we further add the superscripts $^{(n)}$ on $f_V$ and $a^{\mu}f_A$ to highlight their frame dependence due to the presence of magnetization terms. Based on the frame independence of $\mathcal{V}^{\mu}$, we obtain 
\begin{eqnarray}\label{fn_sub}
	&&\delta(q^2-m^2)\Bigg[q^{\mu}\big(f_V^{(n)}-f_V^{(n')}\big)
	\\\nonumber
	&&
	+\hbar\epsilon^{\mu\nu\rho\sigma}\left(\frac{n_{\nu}}{2q\cdot n}-\frac{n'_{\nu}}{2q\cdot n'}\right)\big(\Delta_{\rho}(a_{\sigma}f_A)+F_{\rho\sigma}f_A\big)\Bigg]=0
\end{eqnarray}
up to $\mathcal{O}(\hbar)$ when considering the frame transformation from $n^{\mu}$ to $n'^{\mu}$, where we drop the frame dependence on $a^{\mu}$ and $f_A$ therein since only their frame independent part $\mathcal{O}(\hbar^0)$ contributes. Contracting Eq.~(\ref{fn_sub}) with $n^{\mu}$, one immediately obtain,
\begin{eqnarray}\label{FT_fV}\nonumber
	f^{(n')}_V=f_V^{(n)}+\frac{\hbar\epsilon^{\lambda\nu\rho\sigma}n_{\lambda}n'_{\nu}}{2(q\cdot n)(q\cdot n')}\big(\Delta_{\rho}(a_{\sigma}f_A)+F_{\rho\sigma}f_A\big)
	\\
\end{eqnarray} 
as the modified frame transformation of $f_V$. One may show Eq.~(\ref{FT_fV}) indeed satisfies Eq.~(\ref{fn_sub}) explicitly. By using Eq.~(\ref{FT_fV}) and the Schouten identity~\eqref{Schouten}, it is found
\begin{widetext}
\begin{eqnarray}\nonumber\label{show_fsub}
	&&\delta(q^2-m^2)q^{\mu}\big(f_V^{(n)}-f_V^{(n')}\big)
	\\\nonumber
	&&=\hbar\delta(q^2-m^2)\big(\epsilon^{\mu\nu\rho\sigma}q\cdot nn'_{\nu}+\epsilon^{\lambda\mu\rho\sigma}n_{\lambda}q\cdot n'
	+\epsilon^{\lambda\nu\mu\sigma}n_{\lambda}n'_{\nu}q^{\rho}+\epsilon^{\lambda\nu\rho\mu}n_{\lambda}n'_{\nu}q^{\sigma}
	\big)
	\frac{\big(\Delta_{\rho}(a_{\sigma}f_A)+F_{\rho\sigma}f_A\big)}{2(q\cdot n)(q\cdot n')}
	\\
	&&=\hbar\delta(q^2-m^2)\Bigg[\epsilon^{\mu\nu\rho\sigma}\left(\frac{n'_{\nu}}{2q\cdot n'}-\frac{n_{\nu}}{2q\cdot n}\right)\big(\Delta_{\rho}(a_{\sigma}f_A)+F_{\rho\sigma}f_A\big)
	+\frac{\epsilon^{\lambda\nu\mu\sigma}}{2(q\cdot n)(q\cdot n')}\big(q\cdot\Delta(a_{\sigma}f_A)+F_{\rho\sigma}a^{\sigma}f_A\big)\Bigg],
\end{eqnarray}
\end{widetext}
where we employ $q^{\rho}\Delta_{\sigma}(a_{\rho}f_A)=\Delta_{\sigma}(q\cdot af_A)-F_{\rho\sigma}a^{\rho}f_A$ and 
$q\cdot a=q^2-m^2$ in the computation. Since 
\begin{eqnarray}
	\hbar\delta(q^2-m^2)\big(q\cdot\Delta(a_{\sigma}f_A)+F_{\rho\sigma}a^{\sigma}f_A\big)=\mathcal{O}(\hbar^2)
\end{eqnarray}
according to the AKE and the corresponding term thus can be dropped in Eq.~(\ref{show_fsub}), Eq.~({\ref{fn_sub}}) is indeed satisfied by the modified frame transformation. 
For the frame independence of $\mathcal{A}^{\mu}$, it is straightforward to find
\begin{eqnarray}
	&&a^{({n'})\mu}f_A^{(n')}-a^{(n)\mu}f_A^{(n)}
	\\\nonumber
	&&=\hbar\epsilon^{\mu\nu\alpha\beta}\left(\frac{n_{\beta}}{2(q\cdot n+m)}-\frac{n'_{\beta}}{2(q\cdot n'+m)}\right)q_{\alpha}\Delta_{\nu}f_V
\end{eqnarray}
as the modified frame transformation.
From Eq.~(\ref{FT_fV}), one can make the connection between $n^{\mu}=n^{\mu}(X)$ and the rest frame $n_r^{\mu}=q^{\mu}/m$ through
\begin{eqnarray}
	f^{(n_r)}_V=f_V^{(n)}+\frac{\hbar\epsilon^{\lambda\nu\rho\sigma}n_{\lambda}q_{\nu}}{2(q\cdot n)m^2}\big(\Delta_{\rho}(a_{\sigma}f_A)+F_{\rho\sigma}f_A\big).
\end{eqnarray} 
Nonetheless, in the small-mass region, $f^{(n_r)}_V$ contains a divergent term. 
For $\mathcal{V}^{\mu}$ to be frame invariant, such a divergent term from the modified frame transformation 
should cancel the divergent part of the magnetization current in $n_r^{\mu}$ 
so that the remaining finite part agrees with the magnetization current obtained in $n^{\mu}(X)$. 
One is forced to deal with such a subtle cancellation caused by an inappropriate frame choice 
when $m$ is smaller than the gradient or electromagnetic scales. 
Thanks to our results in the general frame, we may discuss the frame transformation property 
and find how the frame invariance should be realized. 
Even better, we can choose an appropriate frame to avoid such a pathological behavior. 
However, without knowing such a general frame transformation, 
naively working in the rest frame cannot correctly captured the finite quantum effect 
when $m$ is small. 

\bibliography{CKT_mass_paper_PRD.bbl}

\begin{thebibliography}{61}%
\makeatletter
\providecommand \@ifxundefined [1]{%
 \@ifx{#1\undefined}
}%
\providecommand \@ifnum [1]{%
 \ifnum #1\expandafter \@firstoftwo
 \else \expandafter \@secondoftwo
 \fi
}%
\providecommand \@ifx [1]{%
 \ifx #1\expandafter \@firstoftwo
 \else \expandafter \@secondoftwo
 \fi
}%
\providecommand \natexlab [1]{#1}%
\providecommand \enquote  [1]{``#1''}%
\providecommand \bibnamefont  [1]{#1}%
\providecommand \bibfnamefont [1]{#1}%
\providecommand \citenamefont [1]{#1}%
\providecommand \href@noop [0]{\@secondoftwo}%
\providecommand \href [0]{\begingroup \@sanitize@url \@href}%
\providecommand \@href[1]{\@@startlink{#1}\@@href}%
\providecommand \@@href[1]{\endgroup#1\@@endlink}%
\providecommand \@sanitize@url [0]{\catcode `\\12\catcode `\$12\catcode
  `\&12\catcode `\#12\catcode `\^12\catcode `\_12\catcode `\%12\relax}%
\providecommand \@@startlink[1]{}%
\providecommand \@@endlink[0]{}%
\providecommand \url  [0]{\begingroup\@sanitize@url \@url }%
\providecommand \@url [1]{\endgroup\@href {#1}{\urlprefix }}%
\providecommand \urlprefix  [0]{URL }%
\providecommand \Eprint [0]{\href }%
\providecommand \doibase [0]{http://dx.doi.org/}%
\providecommand \selectlanguage [0]{\@gobble}%
\providecommand \bibinfo  [0]{\@secondoftwo}%
\providecommand \bibfield  [0]{\@secondoftwo}%
\providecommand \translation [1]{[#1]}%
\providecommand \BibitemOpen [0]{}%
\providecommand \bibitemStop [0]{}%
\providecommand \bibitemNoStop [0]{.\EOS\space}%
\providecommand \EOS [0]{\spacefactor3000\relax}%
\providecommand \BibitemShut  [1]{\csname bibitem#1\endcsname}%
\let\auto@bib@innerbib\@empty
\bibitem [{\citenamefont {Vilenkin}(1979)}]{Vilenkin:1979ui}%
  \BibitemOpen
  \bibfield  {author} {\bibinfo {author} {\bibfnamefont {A.}~\bibnamefont
  {Vilenkin}},\ }\href {\doibase 10.1103/PhysRevD.20.1807} {\bibfield
  {journal} {\bibinfo  {journal} {Phys. Rev.}\ }\textbf {\bibinfo {volume}
  {D20}},\ \bibinfo {pages} {1807} (\bibinfo {year} {1979})}\BibitemShut
  {NoStop}%
\bibitem [{\citenamefont {Kharzeev}\ \emph {et~al.}(2008)\citenamefont
  {Kharzeev}, \citenamefont {McLerran},\ and\ \citenamefont
  {Warringa}}]{Kharzeev:2007jp}%
  \BibitemOpen
  \bibfield  {author} {\bibinfo {author} {\bibfnamefont {D.~E.}\ \bibnamefont
  {Kharzeev}}, \bibinfo {author} {\bibfnamefont {L.~D.}\ \bibnamefont
  {McLerran}}, \ and\ \bibinfo {author} {\bibfnamefont {H.~J.}\ \bibnamefont
  {Warringa}},\ }\href {\doibase 10.1016/j.nuclphysa.2008.02.298} {\bibfield
  {journal} {\bibinfo  {journal} {Nucl. Phys.}\ }\textbf {\bibinfo {volume}
  {A803}},\ \bibinfo {pages} {227} (\bibinfo {year} {2008})},\ \Eprint
  {http://arxiv.org/abs/0711.0950} {arXiv:0711.0950 [hep-ph]} \BibitemShut
  {NoStop}%
\bibitem [{\citenamefont {Fukushima}\ \emph {et~al.}(2008)\citenamefont
  {Fukushima}, \citenamefont {Kharzeev},\ and\ \citenamefont
  {Warringa}}]{Fukushima:2008xe}%
  \BibitemOpen
  \bibfield  {author} {\bibinfo {author} {\bibfnamefont {K.}~\bibnamefont
  {Fukushima}}, \bibinfo {author} {\bibfnamefont {D.~E.}\ \bibnamefont
  {Kharzeev}}, \ and\ \bibinfo {author} {\bibfnamefont {H.~J.}\ \bibnamefont
  {Warringa}},\ }\href {\doibase 10.1103/PhysRevD.78.074033} {\bibfield
  {journal} {\bibinfo  {journal} {Phys. Rev.}\ }\textbf {\bibinfo {volume}
  {D78}},\ \bibinfo {pages} {074033} (\bibinfo {year} {2008})},\ \Eprint
  {http://arxiv.org/abs/0808.3382} {arXiv:0808.3382 [hep-ph]} \BibitemShut
  {NoStop}%
\bibitem [{\citenamefont {Kharzeev}\ \emph {et~al.}(2016)\citenamefont
  {Kharzeev}, \citenamefont {Liao}, \citenamefont {Voloshin},\ and\
  \citenamefont {Wang}}]{Kharzeev:2015znc}%
  \BibitemOpen
  \bibfield  {author} {\bibinfo {author} {\bibfnamefont {D.~E.}\ \bibnamefont
  {Kharzeev}}, \bibinfo {author} {\bibfnamefont {J.}~\bibnamefont {Liao}},
  \bibinfo {author} {\bibfnamefont {S.~A.}\ \bibnamefont {Voloshin}}, \ and\
  \bibinfo {author} {\bibfnamefont {G.}~\bibnamefont {Wang}},\ }\href {\doibase
  10.1016/j.ppnp.2016.01.001} {\bibfield  {journal} {\bibinfo  {journal} {Prog.
  Part. Nucl. Phys.}\ }\textbf {\bibinfo {volume} {88}},\ \bibinfo {pages} {1}
  (\bibinfo {year} {2016})},\ \Eprint {http://arxiv.org/abs/1511.04050}
  {arXiv:1511.04050 [hep-ph]} \BibitemShut {NoStop}%
\bibitem [{\citenamefont {Hattori}\ and\ \citenamefont
  {Huang}(2017)}]{Hattori:2016emy}%
  \BibitemOpen
  \bibfield  {author} {\bibinfo {author} {\bibfnamefont {K.}~\bibnamefont
  {Hattori}}\ and\ \bibinfo {author} {\bibfnamefont {X.-G.}\ \bibnamefont
  {Huang}},\ }\href {\doibase 10.1007/s41365-016-0178-3} {\bibfield  {journal}
  {\bibinfo  {journal} {Nucl. Sci. Tech.}\ }\textbf {\bibinfo {volume} {28}},\
  \bibinfo {pages} {26} (\bibinfo {year} {2017})},\ \Eprint
  {http://arxiv.org/abs/1609.00747} {arXiv:1609.00747 [nucl-th]} \BibitemShut
  {NoStop}%
\bibitem [{\citenamefont {Li}\ \emph {et~al.}(2016)\citenamefont {Li},
  \citenamefont {Kharzeev}, \citenamefont {Zhang}, \citenamefont {Huang},
  \citenamefont {Pletikosic}, \citenamefont {Fedorov}, \citenamefont {Zhong},
  \citenamefont {Schneeloch}, \citenamefont {Gu},\ and\ \citenamefont
  {Valla}}]{Li:2014bha}%
  \BibitemOpen
  \bibfield  {author} {\bibinfo {author} {\bibfnamefont {Q.}~\bibnamefont
  {Li}}, \bibinfo {author} {\bibfnamefont {D.~E.}\ \bibnamefont {Kharzeev}},
  \bibinfo {author} {\bibfnamefont {C.}~\bibnamefont {Zhang}}, \bibinfo
  {author} {\bibfnamefont {Y.}~\bibnamefont {Huang}}, \bibinfo {author}
  {\bibfnamefont {I.}~\bibnamefont {Pletikosic}}, \bibinfo {author}
  {\bibfnamefont {A.~V.}\ \bibnamefont {Fedorov}}, \bibinfo {author}
  {\bibfnamefont {R.~D.}\ \bibnamefont {Zhong}}, \bibinfo {author}
  {\bibfnamefont {J.~A.}\ \bibnamefont {Schneeloch}}, \bibinfo {author}
  {\bibfnamefont {G.~D.}\ \bibnamefont {Gu}}, \ and\ \bibinfo {author}
  {\bibfnamefont {T.}~\bibnamefont {Valla}},\ }\href {\doibase
  10.1038/nphys3648} {\bibfield  {journal} {\bibinfo  {journal} {Nature Phys.}\
  }\textbf {\bibinfo {volume} {12}},\ \bibinfo {pages} {550} (\bibinfo {year}
  {2016})},\ \Eprint {http://arxiv.org/abs/1412.6543} {arXiv:1412.6543
  [cond-mat.str-el]} \BibitemShut {NoStop}%
\bibitem [{\citenamefont {Yamamoto}(2016)}]{Yamamoto:2015gzz}%
  \BibitemOpen
  \bibfield  {author} {\bibinfo {author} {\bibfnamefont {N.}~\bibnamefont
  {Yamamoto}},\ }\href {\doibase 10.1103/PhysRevD.93.065017} {\bibfield
  {journal} {\bibinfo  {journal} {Phys. Rev.}\ }\textbf {\bibinfo {volume}
  {D93}},\ \bibinfo {pages} {065017} (\bibinfo {year} {2016})},\ \Eprint
  {http://arxiv.org/abs/1511.00933} {arXiv:1511.00933 [astro-ph.HE]}
  \BibitemShut {NoStop}%
\bibitem [{\citenamefont {Masada}\ \emph {et~al.}(2018)\citenamefont {Masada},
  \citenamefont {Kotake}, \citenamefont {Takiwaki},\ and\ \citenamefont
  {Yamamoto}}]{Masada:2018swb}%
  \BibitemOpen
  \bibfield  {author} {\bibinfo {author} {\bibfnamefont {Y.}~\bibnamefont
  {Masada}}, \bibinfo {author} {\bibfnamefont {K.}~\bibnamefont {Kotake}},
  \bibinfo {author} {\bibfnamefont {T.}~\bibnamefont {Takiwaki}}, \ and\
  \bibinfo {author} {\bibfnamefont {N.}~\bibnamefont {Yamamoto}},\ }\href
  {\doibase 10.1103/PhysRevD.98.083018} {\bibfield  {journal} {\bibinfo
  {journal} {Phys. Rev.}\ }\textbf {\bibinfo {volume} {D98}},\ \bibinfo {pages}
  {083018} (\bibinfo {year} {2018})},\ \Eprint
  {http://arxiv.org/abs/1805.10419} {arXiv:1805.10419 [astro-ph.HE]}
  \BibitemShut {NoStop}%
\bibitem [{\citenamefont {Gao}\ \emph {et~al.}(2012)\citenamefont {Gao},
  \citenamefont {Liang}, \citenamefont {Pu}, \citenamefont {Wang},\ and\
  \citenamefont {Wang}}]{Gao:2012ix}%
  \BibitemOpen
  \bibfield  {author} {\bibinfo {author} {\bibfnamefont {J.-H.}\ \bibnamefont
  {Gao}}, \bibinfo {author} {\bibfnamefont {Z.-T.}\ \bibnamefont {Liang}},
  \bibinfo {author} {\bibfnamefont {S.}~\bibnamefont {Pu}}, \bibinfo {author}
  {\bibfnamefont {Q.}~\bibnamefont {Wang}}, \ and\ \bibinfo {author}
  {\bibfnamefont {X.-N.}\ \bibnamefont {Wang}},\ }\href {\doibase
  10.1103/PhysRevLett.109.232301} {\bibfield  {journal} {\bibinfo  {journal}
  {Phys. Rev. Lett.}\ }\textbf {\bibinfo {volume} {109}},\ \bibinfo {pages}
  {232301} (\bibinfo {year} {2012})},\ \Eprint {http://arxiv.org/abs/1203.0725}
  {arXiv:1203.0725 [hep-ph]} \BibitemShut {NoStop}%
\bibitem [{\citenamefont {Son}\ and\ \citenamefont
  {Yamamoto}(2012)}]{Son:2012wh}%
  \BibitemOpen
  \bibfield  {author} {\bibinfo {author} {\bibfnamefont {D.~T.}\ \bibnamefont
  {Son}}\ and\ \bibinfo {author} {\bibfnamefont {N.}~\bibnamefont {Yamamoto}},\
  }\href {\doibase 10.1103/PhysRevLett.109.181602} {\bibfield  {journal}
  {\bibinfo  {journal} {Phys. Rev. Lett.}\ }\textbf {\bibinfo {volume} {109}},\
  \bibinfo {pages} {181602} (\bibinfo {year} {2012})},\ \Eprint
  {http://arxiv.org/abs/1203.2697} {arXiv:1203.2697 [cond-mat.mes-hall]}
  \BibitemShut {NoStop}%
\bibitem [{\citenamefont {Stephanov}\ and\ \citenamefont
  {Yin}(2012)}]{Stephanov:2012ki}%
  \BibitemOpen
  \bibfield  {author} {\bibinfo {author} {\bibfnamefont {M.}~\bibnamefont
  {Stephanov}}\ and\ \bibinfo {author} {\bibfnamefont {Y.}~\bibnamefont
  {Yin}},\ }\href {\doibase 10.1103/PhysRevLett.109.162001} {\bibfield
  {journal} {\bibinfo  {journal} {Phys. Rev. Lett.}\ }\textbf {\bibinfo
  {volume} {109}},\ \bibinfo {pages} {162001} (\bibinfo {year} {2012})},\
  \Eprint {http://arxiv.org/abs/1207.0747} {arXiv:1207.0747 [hep-th]}
  \BibitemShut {NoStop}%
\bibitem [{\citenamefont {Son}\ and\ \citenamefont
  {Yamamoto}(2013)}]{Son:2012zy}%
  \BibitemOpen
  \bibfield  {author} {\bibinfo {author} {\bibfnamefont {D.~T.}\ \bibnamefont
  {Son}}\ and\ \bibinfo {author} {\bibfnamefont {N.}~\bibnamefont {Yamamoto}},\
  }\href {\doibase 10.1103/PhysRevD.87.085016} {\bibfield  {journal} {\bibinfo
  {journal} {Phys. Rev.}\ }\textbf {\bibinfo {volume} {D87}},\ \bibinfo {pages}
  {085016} (\bibinfo {year} {2013})},\ \Eprint {http://arxiv.org/abs/1210.8158}
  {arXiv:1210.8158 [hep-th]} \BibitemShut {NoStop}%
\bibitem [{\citenamefont {Chen}\ \emph {et~al.}(2013)\citenamefont {Chen},
  \citenamefont {Pu}, \citenamefont {Wang},\ and\ \citenamefont
  {Wang}}]{Chen:2012ca}%
  \BibitemOpen
  \bibfield  {author} {\bibinfo {author} {\bibfnamefont {J.-W.}\ \bibnamefont
  {Chen}}, \bibinfo {author} {\bibfnamefont {S.}~\bibnamefont {Pu}}, \bibinfo
  {author} {\bibfnamefont {Q.}~\bibnamefont {Wang}}, \ and\ \bibinfo {author}
  {\bibfnamefont {X.-N.}\ \bibnamefont {Wang}},\ }\href {\doibase
  10.1103/PhysRevLett.110.262301} {\bibfield  {journal} {\bibinfo  {journal}
  {Phys. Rev. Lett.}\ }\textbf {\bibinfo {volume} {110}},\ \bibinfo {pages}
  {262301} (\bibinfo {year} {2013})},\ \Eprint {http://arxiv.org/abs/1210.8312}
  {arXiv:1210.8312 [hep-th]} \BibitemShut {NoStop}%
\bibitem [{\citenamefont {Manuel}\ and\ \citenamefont
  {Torres-Rincon}(2014)}]{Manuel:2013zaa}%
  \BibitemOpen
  \bibfield  {author} {\bibinfo {author} {\bibfnamefont {C.}~\bibnamefont
  {Manuel}}\ and\ \bibinfo {author} {\bibfnamefont {J.~M.}\ \bibnamefont
  {Torres-Rincon}},\ }\href {\doibase 10.1103/PhysRevD.89.096002} {\bibfield
  {journal} {\bibinfo  {journal} {Phys. Rev.}\ }\textbf {\bibinfo {volume}
  {D89}},\ \bibinfo {pages} {096002} (\bibinfo {year} {2014})},\ \Eprint
  {http://arxiv.org/abs/1312.1158} {arXiv:1312.1158 [hep-ph]} \BibitemShut
  {NoStop}%
\bibitem [{\citenamefont {Chen}\ \emph {et~al.}(2014)\citenamefont {Chen},
  \citenamefont {Son}, \citenamefont {Stephanov}, \citenamefont {Yee},\ and\
  \citenamefont {Yin}}]{Chen:2014cla}%
  \BibitemOpen
  \bibfield  {author} {\bibinfo {author} {\bibfnamefont {J.-Y.}\ \bibnamefont
  {Chen}}, \bibinfo {author} {\bibfnamefont {D.~T.}\ \bibnamefont {Son}},
  \bibinfo {author} {\bibfnamefont {M.~A.}\ \bibnamefont {Stephanov}}, \bibinfo
  {author} {\bibfnamefont {H.-U.}\ \bibnamefont {Yee}}, \ and\ \bibinfo
  {author} {\bibfnamefont {Y.}~\bibnamefont {Yin}},\ }\href {\doibase
  10.1103/PhysRevLett.113.182302} {\bibfield  {journal} {\bibinfo  {journal}
  {Phys. Rev. Lett.}\ }\textbf {\bibinfo {volume} {113}},\ \bibinfo {pages}
  {182302} (\bibinfo {year} {2014})},\ \Eprint {http://arxiv.org/abs/1404.5963}
  {arXiv:1404.5963 [hep-th]} \BibitemShut {NoStop}%
\bibitem [{\citenamefont {Chen}\ \emph {et~al.}(2015)\citenamefont {Chen},
  \citenamefont {Son},\ and\ \citenamefont {Stephanov}}]{Chen:2015gta}%
  \BibitemOpen
  \bibfield  {author} {\bibinfo {author} {\bibfnamefont {J.-Y.}\ \bibnamefont
  {Chen}}, \bibinfo {author} {\bibfnamefont {D.~T.}\ \bibnamefont {Son}}, \
  and\ \bibinfo {author} {\bibfnamefont {M.~A.}\ \bibnamefont {Stephanov}},\
  }\href {\doibase 10.1103/PhysRevLett.115.021601} {\bibfield  {journal}
  {\bibinfo  {journal} {Phys. Rev. Lett.}\ }\textbf {\bibinfo {volume} {115}},\
  \bibinfo {pages} {021601} (\bibinfo {year} {2015})},\ \Eprint
  {http://arxiv.org/abs/1502.06966} {arXiv:1502.06966 [hep-th]} \BibitemShut
  {NoStop}%
\bibitem [{\citenamefont {Hidaka}\ \emph {et~al.}(2017)\citenamefont {Hidaka},
  \citenamefont {Pu},\ and\ \citenamefont {Yang}}]{Hidaka:2016yjf}%
  \BibitemOpen
  \bibfield  {author} {\bibinfo {author} {\bibfnamefont {Y.}~\bibnamefont
  {Hidaka}}, \bibinfo {author} {\bibfnamefont {S.}~\bibnamefont {Pu}}, \ and\
  \bibinfo {author} {\bibfnamefont {D.-L.}\ \bibnamefont {Yang}},\ }\href
  {\doibase 10.1103/PhysRevD.95.091901} {\bibfield  {journal} {\bibinfo
  {journal} {Phys. Rev.}\ }\textbf {\bibinfo {volume} {D95}},\ \bibinfo {pages}
  {091901} (\bibinfo {year} {2017})},\ \Eprint
  {http://arxiv.org/abs/1612.04630} {arXiv:1612.04630 [hep-th]} \BibitemShut
  {NoStop}%
\bibitem [{\citenamefont {Hidaka}\ \emph {et~al.}(2018)\citenamefont {Hidaka},
  \citenamefont {Pu},\ and\ \citenamefont {Yang}}]{Hidaka:2017auj}%
  \BibitemOpen
  \bibfield  {author} {\bibinfo {author} {\bibfnamefont {Y.}~\bibnamefont
  {Hidaka}}, \bibinfo {author} {\bibfnamefont {S.}~\bibnamefont {Pu}}, \ and\
  \bibinfo {author} {\bibfnamefont {D.-L.}\ \bibnamefont {Yang}},\ }\href
  {\doibase 10.1103/PhysRevD.97.016004} {\bibfield  {journal} {\bibinfo
  {journal} {Phys. Rev.}\ }\textbf {\bibinfo {volume} {D97}},\ \bibinfo {pages}
  {016004} (\bibinfo {year} {2018})},\ \Eprint
  {http://arxiv.org/abs/1710.00278} {arXiv:1710.00278 [hep-th]} \BibitemShut
  {NoStop}%
\bibitem [{\citenamefont {Hidaka}\ and\ \citenamefont
  {Yang}(2018)}]{Hidaka:2018ekt}%
  \BibitemOpen
  \bibfield  {author} {\bibinfo {author} {\bibfnamefont {Y.}~\bibnamefont
  {Hidaka}}\ and\ \bibinfo {author} {\bibfnamefont {D.-L.}\ \bibnamefont
  {Yang}},\ }\href {\doibase 10.1103/PhysRevD.98.016012} {\bibfield  {journal}
  {\bibinfo  {journal} {Phys. Rev.}\ }\textbf {\bibinfo {volume} {D98}},\
  \bibinfo {pages} {016012} (\bibinfo {year} {2018})},\ \Eprint
  {http://arxiv.org/abs/1801.08253} {arXiv:1801.08253 [hep-th]} \BibitemShut
  {NoStop}%
\bibitem [{\citenamefont {Mueller}\ and\ \citenamefont
  {Venugopalan}(2017{\natexlab{a}})}]{Mueller:2017arw}%
  \BibitemOpen
  \bibfield  {author} {\bibinfo {author} {\bibfnamefont {N.}~\bibnamefont
  {Mueller}}\ and\ \bibinfo {author} {\bibfnamefont {R.}~\bibnamefont
  {Venugopalan}},\ }\href {\doibase 10.1103/PhysRevD.96.016023} {\bibfield
  {journal} {\bibinfo  {journal} {Phys. Rev.}\ }\textbf {\bibinfo {volume}
  {D96}},\ \bibinfo {pages} {016023} (\bibinfo {year} {2017}{\natexlab{a}})},\
  \Eprint {http://arxiv.org/abs/1702.01233} {arXiv:1702.01233 [hep-ph]}
  \BibitemShut {NoStop}%
\bibitem [{\citenamefont {Mueller}\ and\ \citenamefont
  {Venugopalan}(2017{\natexlab{b}})}]{Mueller:2017lzw}%
  \BibitemOpen
  \bibfield  {author} {\bibinfo {author} {\bibfnamefont {N.}~\bibnamefont
  {Mueller}}\ and\ \bibinfo {author} {\bibfnamefont {R.}~\bibnamefont
  {Venugopalan}},\ }\href@noop {} {\  (\bibinfo {year} {2017}{\natexlab{b}})},\
  \Eprint {http://arxiv.org/abs/1701.03331} {arXiv:1701.03331 [hep-ph]}
  \BibitemShut {NoStop}%
\bibitem [{\citenamefont {Huang}\ \emph {et~al.}(2018)\citenamefont {Huang},
  \citenamefont {Shi}, \citenamefont {Jiang}, \citenamefont {Liao},\ and\
  \citenamefont {Zhuang}}]{Huang:2018wdl}%
  \BibitemOpen
  \bibfield  {author} {\bibinfo {author} {\bibfnamefont {A.}~\bibnamefont
  {Huang}}, \bibinfo {author} {\bibfnamefont {S.}~\bibnamefont {Shi}}, \bibinfo
  {author} {\bibfnamefont {Y.}~\bibnamefont {Jiang}}, \bibinfo {author}
  {\bibfnamefont {J.}~\bibnamefont {Liao}}, \ and\ \bibinfo {author}
  {\bibfnamefont {P.}~\bibnamefont {Zhuang}},\ }\href {\doibase
  10.1103/PhysRevD.98.036010} {\bibfield  {journal} {\bibinfo  {journal} {Phys.
  Rev.}\ }\textbf {\bibinfo {volume} {D98}},\ \bibinfo {pages} {036010}
  (\bibinfo {year} {2018})},\ \Eprint {http://arxiv.org/abs/1801.03640}
  {arXiv:1801.03640 [hep-th]} \BibitemShut {NoStop}%
\bibitem [{\citenamefont {Carignano}\ \emph {et~al.}(2018)\citenamefont
  {Carignano}, \citenamefont {Manuel},\ and\ \citenamefont
  {Torres-Rincon}}]{Carignano:2018gqt}%
  \BibitemOpen
  \bibfield  {author} {\bibinfo {author} {\bibfnamefont {S.}~\bibnamefont
  {Carignano}}, \bibinfo {author} {\bibfnamefont {C.}~\bibnamefont {Manuel}}, \
  and\ \bibinfo {author} {\bibfnamefont {J.~M.}\ \bibnamefont
  {Torres-Rincon}},\ }\href {\doibase 10.1103/PhysRevD.98.076005} {\bibfield
  {journal} {\bibinfo  {journal} {Phys. Rev.}\ }\textbf {\bibinfo {volume}
  {D98}},\ \bibinfo {pages} {076005} (\bibinfo {year} {2018})},\ \Eprint
  {http://arxiv.org/abs/1806.01684} {arXiv:1806.01684 [hep-ph]} \BibitemShut
  {NoStop}%
\bibitem [{\citenamefont {Dayi}\ and\ \citenamefont
  {Kilinçarslan}(2018)}]{Dayi:2018xdy}%
  \BibitemOpen
  \bibfield  {author} {\bibinfo {author} {\bibfnamefont {m.~F.}\ \bibnamefont
  {Dayi}}\ and\ \bibinfo {author} {\bibfnamefont {E.}~\bibnamefont
  {Kilinçarslan}},\ }\href {\doibase 10.1103/PhysRevD.98.081701} {\bibfield
  {journal} {\bibinfo  {journal} {Phys. Rev.}\ }\textbf {\bibinfo {volume}
  {D98}},\ \bibinfo {pages} {081701} (\bibinfo {year} {2018})},\ \Eprint
  {http://arxiv.org/abs/1807.05912} {arXiv:1807.05912 [hep-th]} \BibitemShut
  {NoStop}%
\bibitem [{\citenamefont {Liu}\ \emph {et~al.}(2018)\citenamefont {Liu},
  \citenamefont {Gao}, \citenamefont {Mameda},\ and\ \citenamefont
  {Huang}}]{Liu:2018xip}%
  \BibitemOpen
  \bibfield  {author} {\bibinfo {author} {\bibfnamefont {Y.-C.}\ \bibnamefont
  {Liu}}, \bibinfo {author} {\bibfnamefont {L.-L.}\ \bibnamefont {Gao}},
  \bibinfo {author} {\bibfnamefont {K.}~\bibnamefont {Mameda}}, \ and\ \bibinfo
  {author} {\bibfnamefont {X.-G.}\ \bibnamefont {Huang}},\ }\href@noop {} {\
  (\bibinfo {year} {2018})},\ \Eprint {http://arxiv.org/abs/1812.10127}
  {arXiv:1812.10127 [hep-th]} \BibitemShut {NoStop}%
\bibitem [{\citenamefont {Lin}\ and\ \citenamefont
  {Shukla}(2019)}]{Lin:2019ytz}%
  \BibitemOpen
  \bibfield  {author} {\bibinfo {author} {\bibfnamefont {S.}~\bibnamefont
  {Lin}}\ and\ \bibinfo {author} {\bibfnamefont {A.}~\bibnamefont {Shukla}},\
  }\href@noop {} {\  (\bibinfo {year} {2019})},\ \Eprint
  {http://arxiv.org/abs/1901.01528} {arXiv:1901.01528 [hep-ph]} \BibitemShut
  {NoStop}%
\bibitem [{\citenamefont {Heinz}(1983)}]{Heinz:1983nx}%
  \BibitemOpen
  \bibfield  {author} {\bibinfo {author} {\bibfnamefont {U.~W.}\ \bibnamefont
  {Heinz}},\ }\href {\doibase 10.1103/PhysRevLett.51.351} {\bibfield  {journal}
  {\bibinfo  {journal} {Phys. Rev. Lett.}\ }\textbf {\bibinfo {volume} {51}},\
  \bibinfo {pages} {351} (\bibinfo {year} {1983})}\BibitemShut {NoStop}%
\bibitem [{\citenamefont {Elze}\ \emph {et~al.}(1986)\citenamefont {Elze},
  \citenamefont {Gyulassy},\ and\ \citenamefont {Vasak}}]{Elze:1986hq}%
  \BibitemOpen
  \bibfield  {author} {\bibinfo {author} {\bibfnamefont {H.~T.}\ \bibnamefont
  {Elze}}, \bibinfo {author} {\bibfnamefont {M.}~\bibnamefont {Gyulassy}}, \
  and\ \bibinfo {author} {\bibfnamefont {D.}~\bibnamefont {Vasak}},\ }\href
  {\doibase 10.1016/0370-2693(86)90778-1} {\bibfield  {journal} {\bibinfo
  {journal} {Phys. Lett.}\ }\textbf {\bibinfo {volume} {B177}},\ \bibinfo
  {pages} {402} (\bibinfo {year} {1986})}\BibitemShut {NoStop}%
\bibitem [{\citenamefont {Vasak}\ \emph {et~al.}(1987)\citenamefont {Vasak},
  \citenamefont {Gyulassy},\ and\ \citenamefont {Elze}}]{Vasak:1987um}%
  \BibitemOpen
  \bibfield  {author} {\bibinfo {author} {\bibfnamefont {D.}~\bibnamefont
  {Vasak}}, \bibinfo {author} {\bibfnamefont {M.}~\bibnamefont {Gyulassy}}, \
  and\ \bibinfo {author} {\bibfnamefont {H.~T.}\ \bibnamefont {Elze}},\ }\href
  {\doibase 10.1016/0003-4916(87)90169-2} {\bibfield  {journal} {\bibinfo
  {journal} {Annals Phys.}\ }\textbf {\bibinfo {volume} {173}},\ \bibinfo
  {pages} {462} (\bibinfo {year} {1987})}\BibitemShut {NoStop}%
\bibitem [{\citenamefont {Blaizot}\ and\ \citenamefont
  {Iancu}(2002)}]{Blaizot:2001nr}%
  \BibitemOpen
  \bibfield  {author} {\bibinfo {author} {\bibfnamefont {J.-P.}\ \bibnamefont
  {Blaizot}}\ and\ \bibinfo {author} {\bibfnamefont {E.}~\bibnamefont
  {Iancu}},\ }\href {\doibase 10.1016/S0370-1573(01)00061-8} {\bibfield
  {journal} {\bibinfo  {journal} {Phys. Rept.}\ }\textbf {\bibinfo {volume}
  {359}},\ \bibinfo {pages} {355} (\bibinfo {year} {2002})},\ \Eprint
  {http://arxiv.org/abs/hep-ph/0101103} {arXiv:hep-ph/0101103 [hep-ph]}
  \BibitemShut {NoStop}%
\bibitem [{\citenamefont {Mueller}\ and\ \citenamefont
  {Venugopalan}(2019)}]{Mueller:2019gjj}%
  \BibitemOpen
  \bibfield  {author} {\bibinfo {author} {\bibfnamefont {N.}~\bibnamefont
  {Mueller}}\ and\ \bibinfo {author} {\bibfnamefont {R.}~\bibnamefont
  {Venugopalan}},\ }\href@noop {} {\  (\bibinfo {year} {2019})},\ \Eprint
  {http://arxiv.org/abs/1901.10492} {arXiv:1901.10492 [hep-th]} \BibitemShut
  {NoStop}%
\bibitem [{\citenamefont {Adamczyk}\ \emph {et~al.}(2017)\citenamefont
  {Adamczyk} \emph {et~al.}}]{STAR:2017ckg}%
  \BibitemOpen
  \bibfield  {author} {\bibinfo {author} {\bibfnamefont {L.}~\bibnamefont
  {Adamczyk}} \emph {et~al.} (\bibinfo {collaboration} {STAR}),\ }\href
  {\doibase 10.1038/nature23004} {\bibfield  {journal} {\bibinfo  {journal}
  {Nature}\ }\textbf {\bibinfo {volume} {548}},\ \bibinfo {pages} {62}
  (\bibinfo {year} {2017})},\ \Eprint {http://arxiv.org/abs/1701.06657}
  {arXiv:1701.06657 [nucl-ex]} \BibitemShut {NoStop}%
\bibitem [{\citenamefont {Adam}\ \emph {et~al.}(2018)\citenamefont {Adam} \emph
  {et~al.}}]{Adam:2018ivw}%
  \BibitemOpen
  \bibfield  {author} {\bibinfo {author} {\bibfnamefont {J.}~\bibnamefont
  {Adam}} \emph {et~al.} (\bibinfo {collaboration} {STAR}),\ }\href@noop {} {\
  (\bibinfo {year} {2018})},\ \Eprint {http://arxiv.org/abs/1805.04400}
  {arXiv:1805.04400 [nucl-ex]} \BibitemShut {NoStop}%
\bibitem [{\citenamefont {Liang}\ and\ \citenamefont
  {Wang}(2005)}]{Liang:2004ph}%
  \BibitemOpen
  \bibfield  {author} {\bibinfo {author} {\bibfnamefont {Z.-T.}\ \bibnamefont
  {Liang}}\ and\ \bibinfo {author} {\bibfnamefont {X.-N.}\ \bibnamefont
  {Wang}},\ }\href {\doibase 10.1103/PhysRevLett.94.102301,
  10.1103/PhysRevLett.96.039901} {\bibfield  {journal} {\bibinfo  {journal}
  {Phys. Rev. Lett.}\ }\textbf {\bibinfo {volume} {94}},\ \bibinfo {pages}
  {102301} (\bibinfo {year} {2005})},\ \bibinfo {note} {[Erratum: Phys. Rev.
  Lett.96,039901(2006)]},\ \Eprint {http://arxiv.org/abs/nucl-th/0410079}
  {arXiv:nucl-th/0410079 [nucl-th]} \BibitemShut {NoStop}%
\bibitem [{\citenamefont {Becattini}\ \emph
  {et~al.}(2013{\natexlab{a}})\citenamefont {Becattini}, \citenamefont
  {Csernai},\ and\ \citenamefont {Wang}}]{Becattini:2013vja}%
  \BibitemOpen
  \bibfield  {author} {\bibinfo {author} {\bibfnamefont {F.}~\bibnamefont
  {Becattini}}, \bibinfo {author} {\bibfnamefont {L.}~\bibnamefont {Csernai}},
  \ and\ \bibinfo {author} {\bibfnamefont {D.~J.}\ \bibnamefont {Wang}},\
  }\href {\doibase 10.1103/PhysRevC.93.069901, 10.1103/PhysRevC.88.034905}
  {\bibfield  {journal} {\bibinfo  {journal} {Phys. Rev.}\ }\textbf {\bibinfo
  {volume} {C88}},\ \bibinfo {pages} {034905} (\bibinfo {year}
  {2013}{\natexlab{a}})},\ \bibinfo {note} {[Erratum: Phys.
  Rev.C93,no.6,069901(2016)]},\ \Eprint {http://arxiv.org/abs/1304.4427}
  {arXiv:1304.4427 [nucl-th]} \BibitemShut {NoStop}%
\bibitem [{\citenamefont {Becattini}\ \emph
  {et~al.}(2013{\natexlab{b}})\citenamefont {Becattini}, \citenamefont
  {Chandra}, \citenamefont {Del~Zanna},\ and\ \citenamefont
  {Grossi}}]{Becattini2013a}%
  \BibitemOpen
  \bibfield  {author} {\bibinfo {author} {\bibfnamefont {F.}~\bibnamefont
  {Becattini}}, \bibinfo {author} {\bibfnamefont {V.}~\bibnamefont {Chandra}},
  \bibinfo {author} {\bibfnamefont {L.}~\bibnamefont {Del~Zanna}}, \ and\
  \bibinfo {author} {\bibfnamefont {E.}~\bibnamefont {Grossi}},\ }\href
  {\doibase 10.1016/j.aop.2013.07.004} {\bibfield  {journal} {\bibinfo
  {journal} {Annals Phys.}\ }\textbf {\bibinfo {volume} {338}},\ \bibinfo
  {pages} {32} (\bibinfo {year} {2013}{\natexlab{b}})},\ \Eprint
  {http://arxiv.org/abs/1303.3431} {arXiv:1303.3431 [nucl-th]} \BibitemShut
  {NoStop}%
\bibitem [{\citenamefont {Fang}\ \emph {et~al.}(2016)\citenamefont {Fang},
  \citenamefont {Pang}, \citenamefont {Wang},\ and\ \citenamefont
  {Wang}}]{Fang:2016vpj}%
  \BibitemOpen
  \bibfield  {author} {\bibinfo {author} {\bibfnamefont {R.-h.}\ \bibnamefont
  {Fang}}, \bibinfo {author} {\bibfnamefont {L.-g.}\ \bibnamefont {Pang}},
  \bibinfo {author} {\bibfnamefont {Q.}~\bibnamefont {Wang}}, \ and\ \bibinfo
  {author} {\bibfnamefont {X.-n.}\ \bibnamefont {Wang}},\ }\href@noop {} {\
  (\bibinfo {year} {2016})},\ \Eprint {http://arxiv.org/abs/1604.04036}
  {arXiv:1604.04036 [nucl-th]} \BibitemShut {NoStop}%
\bibitem [{\citenamefont {Florkowski}\ \emph
  {et~al.}(2018{\natexlab{a}})\citenamefont {Florkowski}, \citenamefont
  {Friman}, \citenamefont {Jaiswal},\ and\ \citenamefont
  {Speranza}}]{Florkowski:2017ruc}%
  \BibitemOpen
  \bibfield  {author} {\bibinfo {author} {\bibfnamefont {W.}~\bibnamefont
  {Florkowski}}, \bibinfo {author} {\bibfnamefont {B.}~\bibnamefont {Friman}},
  \bibinfo {author} {\bibfnamefont {A.}~\bibnamefont {Jaiswal}}, \ and\
  \bibinfo {author} {\bibfnamefont {E.}~\bibnamefont {Speranza}},\ }\href
  {\doibase 10.1103/PhysRevC.97.041901} {\bibfield  {journal} {\bibinfo
  {journal} {Phys. Rev.}\ }\textbf {\bibinfo {volume} {C97}},\ \bibinfo {pages}
  {041901} (\bibinfo {year} {2018}{\natexlab{a}})},\ \Eprint
  {http://arxiv.org/abs/1705.00587} {arXiv:1705.00587 [nucl-th]} \BibitemShut
  {NoStop}%
\bibitem [{\citenamefont {Yang}(2018)}]{Yang:2018lew}%
  \BibitemOpen
  \bibfield  {author} {\bibinfo {author} {\bibfnamefont {D.-L.}\ \bibnamefont
  {Yang}},\ }\href {\doibase 10.1103/PhysRevD.98.076019} {\bibfield  {journal}
  {\bibinfo  {journal} {Phys. Rev.}\ }\textbf {\bibinfo {volume} {D98}},\
  \bibinfo {pages} {076019} (\bibinfo {year} {2018})},\ \Eprint
  {http://arxiv.org/abs/1807.02395} {arXiv:1807.02395 [nucl-th]} \BibitemShut
  {NoStop}%
\bibitem [{\citenamefont {Fukushima}\ \emph {et~al.}(2018)\citenamefont
  {Fukushima}, \citenamefont {Pu},\ and\ \citenamefont
  {Qiu}}]{Fukushima:2018osn}%
  \BibitemOpen
  \bibfield  {author} {\bibinfo {author} {\bibfnamefont {K.}~\bibnamefont
  {Fukushima}}, \bibinfo {author} {\bibfnamefont {S.}~\bibnamefont {Pu}}, \
  and\ \bibinfo {author} {\bibfnamefont {Z.}~\bibnamefont {Qiu}},\ }\href@noop
  {} {\  (\bibinfo {year} {2018})},\ \Eprint {http://arxiv.org/abs/1808.08016}
  {arXiv:1808.08016 [hep-ph]} \BibitemShut {NoStop}%
\bibitem [{\citenamefont {Florkowski}\ and\ \citenamefont
  {Ryblewski}(2018)}]{Florkowski:2018fap}%
  \BibitemOpen
  \bibfield  {author} {\bibinfo {author} {\bibfnamefont {W.}~\bibnamefont
  {Florkowski}}\ and\ \bibinfo {author} {\bibfnamefont {R.}~\bibnamefont
  {Ryblewski}},\ }\href@noop {} {\  (\bibinfo {year} {2018})},\ \Eprint
  {http://arxiv.org/abs/1811.04409} {arXiv:1811.04409 [nucl-th]} \BibitemShut
  {NoStop}%
\bibitem [{\citenamefont {Florkowski}\ \emph
  {et~al.}(2018{\natexlab{b}})\citenamefont {Florkowski}, \citenamefont
  {Kumar},\ and\ \citenamefont {Ryblewski}}]{Florkowski:2018ahw}%
  \BibitemOpen
  \bibfield  {author} {\bibinfo {author} {\bibfnamefont {W.}~\bibnamefont
  {Florkowski}}, \bibinfo {author} {\bibfnamefont {A.}~\bibnamefont {Kumar}}, \
  and\ \bibinfo {author} {\bibfnamefont {R.}~\bibnamefont {Ryblewski}},\ }\href
  {\doibase 10.1103/PhysRevC.98.044906} {\bibfield  {journal} {\bibinfo
  {journal} {Phys. Rev.}\ }\textbf {\bibinfo {volume} {C98}},\ \bibinfo {pages}
  {044906} (\bibinfo {year} {2018}{\natexlab{b}})},\ \Eprint
  {http://arxiv.org/abs/1806.02616} {arXiv:1806.02616 [hep-ph]} \BibitemShut
  {NoStop}%
\bibitem [{\citenamefont {Hattori}\ \emph {et~al.}(2019)\citenamefont
  {Hattori}, \citenamefont {Hongo}, \citenamefont {Huang}, \citenamefont
  {Matsuo},\ and\ \citenamefont {Taya}}]{Hattori:2019lfp}%
  \BibitemOpen
  \bibfield  {author} {\bibinfo {author} {\bibfnamefont {K.}~\bibnamefont
  {Hattori}}, \bibinfo {author} {\bibfnamefont {M.}~\bibnamefont {Hongo}},
  \bibinfo {author} {\bibfnamefont {X.-G.}\ \bibnamefont {Huang}}, \bibinfo
  {author} {\bibfnamefont {M.}~\bibnamefont {Matsuo}}, \ and\ \bibinfo {author}
  {\bibfnamefont {H.}~\bibnamefont {Taya}},\ }\href@noop {} {\  (\bibinfo
  {year} {2019})},\ \Eprint {http://arxiv.org/abs/1901.06615} {arXiv:1901.06615
  [hep-th]} \BibitemShut {NoStop}%
\bibitem [{\citenamefont {Gorbar}\ \emph {et~al.}(2013)\citenamefont {Gorbar},
  \citenamefont {Miransky}, \citenamefont {Shovkovy},\ and\ \citenamefont
  {Wang}}]{Gorbar2013}%
  \BibitemOpen
  \bibfield  {author} {\bibinfo {author} {\bibfnamefont {E.}~\bibnamefont
  {Gorbar}}, \bibinfo {author} {\bibfnamefont {V.}~\bibnamefont {Miransky}},
  \bibinfo {author} {\bibfnamefont {I.}~\bibnamefont {Shovkovy}}, \ and\
  \bibinfo {author} {\bibfnamefont {X.}~\bibnamefont {Wang}},\ }\href {\doibase
  10.1103/PhysRevD.88.025025} {\bibfield  {journal} {\bibinfo  {journal} {Phys.
  Rev.}\ }\textbf {\bibinfo {volume} {D88}},\ \bibinfo {pages} {025025}
  (\bibinfo {year} {2013})},\ \Eprint {http://arxiv.org/abs/1304.4606}
  {arXiv:1304.4606 [hep-ph]} \BibitemShut {NoStop}%
\bibitem [{\citenamefont {Buzzegoli}\ \emph {et~al.}(2017)\citenamefont
  {Buzzegoli}, \citenamefont {Grossi},\ and\ \citenamefont
  {Becattini}}]{Buzzegoli:2017cqy}%
  \BibitemOpen
  \bibfield  {author} {\bibinfo {author} {\bibfnamefont {M.}~\bibnamefont
  {Buzzegoli}}, \bibinfo {author} {\bibfnamefont {E.}~\bibnamefont {Grossi}}, \
  and\ \bibinfo {author} {\bibfnamefont {F.}~\bibnamefont {Becattini}},\ }\href
  {\doibase 10.1007/JHEP07(2018)119, 10.1007/JHEP10(2017)091} {\bibfield
  {journal} {\bibinfo  {journal} {JHEP}\ }\textbf {\bibinfo {volume} {10}},\
  \bibinfo {pages} {091} (\bibinfo {year} {2017})},\ \bibinfo {note} {[Erratum:
  JHEP07,119(2018)]},\ \Eprint {http://arxiv.org/abs/1704.02808}
  {arXiv:1704.02808 [hep-th]} \BibitemShut {NoStop}%
\bibitem [{\citenamefont {Lin}\ and\ \citenamefont {Yang}(2018)}]{Lin:2018aon}%
  \BibitemOpen
  \bibfield  {author} {\bibinfo {author} {\bibfnamefont {S.}~\bibnamefont
  {Lin}}\ and\ \bibinfo {author} {\bibfnamefont {L.}~\bibnamefont {Yang}},\
  }\href {\doibase 10.1103/PhysRevD.98.114022} {\bibfield  {journal} {\bibinfo
  {journal} {Phys. Rev.}\ }\textbf {\bibinfo {volume} {D98}},\ \bibinfo {pages}
  {114022} (\bibinfo {year} {2018})},\ \Eprint
  {http://arxiv.org/abs/1810.02979} {arXiv:1810.02979 [nucl-th]} \BibitemShut
  {NoStop}%
\bibitem [{\citenamefont {Akamatsu}\ and\ \citenamefont
  {Yamamoto}(2013)}]{Akamatsu:2013pjd}%
  \BibitemOpen
  \bibfield  {author} {\bibinfo {author} {\bibfnamefont {Y.}~\bibnamefont
  {Akamatsu}}\ and\ \bibinfo {author} {\bibfnamefont {N.}~\bibnamefont
  {Yamamoto}},\ }\href {\doibase 10.1103/PhysRevLett.111.052002} {\bibfield
  {journal} {\bibinfo  {journal} {Phys. Rev. Lett.}\ }\textbf {\bibinfo
  {volume} {111}},\ \bibinfo {pages} {052002} (\bibinfo {year} {2013})},\
  \Eprint {http://arxiv.org/abs/1302.2125} {arXiv:1302.2125 [nucl-th]}
  \BibitemShut {NoStop}%
\bibitem [{\citenamefont {Ohnishi}\ and\ \citenamefont
  {Yamamoto}(2014)}]{Ohnishi:2014uea}%
  \BibitemOpen
  \bibfield  {author} {\bibinfo {author} {\bibfnamefont {A.}~\bibnamefont
  {Ohnishi}}\ and\ \bibinfo {author} {\bibfnamefont {N.}~\bibnamefont
  {Yamamoto}},\ }\href@noop {} {\  (\bibinfo {year} {2014})},\ \Eprint
  {http://arxiv.org/abs/1402.4760} {arXiv:1402.4760 [astro-ph.HE]} \BibitemShut
  {NoStop}%
\bibitem [{\citenamefont {Grabowska}\ \emph {et~al.}(2015)\citenamefont
  {Grabowska}, \citenamefont {Kaplan},\ and\ \citenamefont
  {Reddy}}]{Grabowska:2014efa}%
  \BibitemOpen
  \bibfield  {author} {\bibinfo {author} {\bibfnamefont {D.}~\bibnamefont
  {Grabowska}}, \bibinfo {author} {\bibfnamefont {D.~B.}\ \bibnamefont
  {Kaplan}}, \ and\ \bibinfo {author} {\bibfnamefont {S.}~\bibnamefont
  {Reddy}},\ }\href {\doibase 10.1103/PhysRevD.91.085035} {\bibfield  {journal}
  {\bibinfo  {journal} {Phys. Rev.}\ }\textbf {\bibinfo {volume} {D91}},\
  \bibinfo {pages} {085035} (\bibinfo {year} {2015})},\ \Eprint
  {http://arxiv.org/abs/1409.3602} {arXiv:1409.3602 [hep-ph]} \BibitemShut
  {NoStop}%
\bibitem [{\citenamefont {Weickgenannt}\ \emph {et~al.}(2019)\citenamefont
  {Weickgenannt}, \citenamefont {Sheng}, \citenamefont {Speranza},
  \citenamefont {Wang},\ and\ \citenamefont {Rischke}}]{Weickgenannt:2019dks}%
  \BibitemOpen
  \bibfield  {author} {\bibinfo {author} {\bibfnamefont {N.}~\bibnamefont
  {Weickgenannt}}, \bibinfo {author} {\bibfnamefont {X.-l.}\ \bibnamefont
  {Sheng}}, \bibinfo {author} {\bibfnamefont {E.}~\bibnamefont {Speranza}},
  \bibinfo {author} {\bibfnamefont {Q.}~\bibnamefont {Wang}}, \ and\ \bibinfo
  {author} {\bibfnamefont {D.~H.}\ \bibnamefont {Rischke}},\ }\href@noop {} {\
  (\bibinfo {year} {2019})},\ \Eprint {http://arxiv.org/abs/1902.06513}
  {arXiv:1902.06513 [hep-ph]} \BibitemShut {NoStop}%
\bibitem [{\citenamefont {Gao}\ and\ \citenamefont
  {Liang}(2019)}]{Gao:2019znl}%
  \BibitemOpen
  \bibfield  {author} {\bibinfo {author} {\bibfnamefont {J.-H.}\ \bibnamefont
  {Gao}}\ and\ \bibinfo {author} {\bibfnamefont {Z.-T.}\ \bibnamefont
  {Liang}},\ }\href@noop {} {\  (\bibinfo {year} {2019})},\ \Eprint
  {http://arxiv.org/abs/1902.06510} {arXiv:1902.06510 [hep-ph]} \BibitemShut
  {NoStop}%
\bibitem [{\citenamefont {Ochs}\ and\ \citenamefont
  {Heinz}(1998)}]{Ochs:1998qj}%
  \BibitemOpen
  \bibfield  {author} {\bibinfo {author} {\bibfnamefont {S.}~\bibnamefont
  {Ochs}}\ and\ \bibinfo {author} {\bibfnamefont {U.~W.}\ \bibnamefont
  {Heinz}},\ }\href {\doibase 10.1006/aphy.1998.5796} {\bibfield  {journal}
  {\bibinfo  {journal} {Annals Phys.}\ }\textbf {\bibinfo {volume} {266}},\
  \bibinfo {pages} {351} (\bibinfo {year} {1998})},\ \Eprint
  {http://arxiv.org/abs/hep-th/9806118} {arXiv:hep-th/9806118 [hep-th]}
  \BibitemShut {NoStop}%
\bibitem [{\citenamefont {Bargmann}\ \emph {et~al.}(1959)\citenamefont
  {Bargmann}, \citenamefont {Michel},\ and\ \citenamefont
  {Telegdi}}]{BLT_spin}%
  \BibitemOpen
  \bibfield  {author} {\bibinfo {author} {\bibfnamefont {V.}~\bibnamefont
  {Bargmann}}, \bibinfo {author} {\bibfnamefont {L.}~\bibnamefont {Michel}}, \
  and\ \bibinfo {author} {\bibfnamefont {V.~L.}\ \bibnamefont {Telegdi}},\
  }\href {\doibase 10.1103/PhysRevLett.2.435} {\bibfield  {journal} {\bibinfo
  {journal} {Phys. Rev. Lett.}\ }\textbf {\bibinfo {volume} {2}},\ \bibinfo
  {pages} {435} (\bibinfo {year} {1959})}\BibitemShut {NoStop}%
\bibitem [{Note1()}]{Note1}%
  \BibitemOpen
  \bibinfo {note} {Similar to the massless case \cite {Hidaka:2016yjf}, the
  $\protect \mathcal {O}(\hbar ^1)$ terms proportional to $q^{\mu }\delta
  (q^2-m^2)$ can be absorbed into the distribution functions.}\BibitemShut
  {Stop}%
\bibitem [{\citenamefont {Peskin}\ and\ \citenamefont
  {Schroeder}(1995)}]{Peskin}%
  \BibitemOpen
  \bibfield  {author} {\bibinfo {author} {\bibfnamefont {M.}~\bibnamefont
  {Peskin}}\ and\ \bibinfo {author} {\bibfnamefont {D.}~\bibnamefont
  {Schroeder}},\ }\href@noop {} {\emph {\bibinfo {title} {An Introduction to
  Quantum Field Theory}}}\ (\bibinfo {year} {1995})\BibitemShut {NoStop}%
\bibitem [{Note2()}]{Note2}%
  \BibitemOpen
  \bibinfo {note} {To be more precise, for both positive and negative $q\cdot
  n$, we should write $2S^{\mu \nu }_{m(n)}=\protect \mathaccentV
  {bar}016{\epsilon }(q\cdot n)\epsilon ^{\mu \nu \alpha \beta }q_{\alpha
  }n_{\beta }/(|q\cdot n|+m)$, where $\protect \mathaccentV {bar}016{\epsilon
  }(q\cdot n)$ denotes the sign for $q\cdot n$.}\BibitemShut {Stop}%
\bibitem [{Note3()}]{Note3}%
  \BibitemOpen
  \bibinfo {note} {More precisely, one has to include both fermions and
  anti-fermions, and put an overall sign function $\protect \mathaccentV
  {bar}016{\epsilon }(q\cdot n)$ in front of the Wigner function.}\BibitemShut
  {Stop}%
\bibitem [{\citenamefont {Prokhorov}\ and\ \citenamefont
  {Teryaev}(2018)}]{Prokhorov:2017atp}%
  \BibitemOpen
  \bibfield  {author} {\bibinfo {author} {\bibfnamefont {G.}~\bibnamefont
  {Prokhorov}}\ and\ \bibinfo {author} {\bibfnamefont {O.}~\bibnamefont
  {Teryaev}},\ }\href {\doibase 10.1103/PhysRevD.97.076013} {\bibfield
  {journal} {\bibinfo  {journal} {Phys. Rev.}\ }\textbf {\bibinfo {volume}
  {D97}},\ \bibinfo {pages} {076013} (\bibinfo {year} {2018})},\ \Eprint
  {http://arxiv.org/abs/1707.02491} {arXiv:1707.02491 [hep-th]} \BibitemShut
  {NoStop}%
\bibitem [{\citenamefont {Prokhorov}\ \emph {et~al.}(2018)\citenamefont
  {Prokhorov}, \citenamefont {Teryaev},\ and\ \citenamefont
  {Zakharov}}]{Prokhorov:2018qhq}%
  \BibitemOpen
  \bibfield  {author} {\bibinfo {author} {\bibfnamefont {G.}~\bibnamefont
  {Prokhorov}}, \bibinfo {author} {\bibfnamefont {O.}~\bibnamefont {Teryaev}},
  \ and\ \bibinfo {author} {\bibfnamefont {V.}~\bibnamefont {Zakharov}},\
  }\href {\doibase 10.1103/PhysRevD.98.071901} {\bibfield  {journal} {\bibinfo
  {journal} {Phys. Rev.}\ }\textbf {\bibinfo {volume} {D98}},\ \bibinfo {pages}
  {071901} (\bibinfo {year} {2018})},\ \Eprint
  {http://arxiv.org/abs/1805.12029} {arXiv:1805.12029 [hep-th]} \BibitemShut
  {NoStop}%
\bibitem [{\citenamefont {Fukushima}\ \emph {et~al.}(2010)\citenamefont
  {Fukushima}, \citenamefont {Kharzeev},\ and\ \citenamefont
  {Warringa}}]{Fukushima:2009ft}%
  \BibitemOpen
  \bibfield  {author} {\bibinfo {author} {\bibfnamefont {K.}~\bibnamefont
  {Fukushima}}, \bibinfo {author} {\bibfnamefont {D.~E.}\ \bibnamefont
  {Kharzeev}}, \ and\ \bibinfo {author} {\bibfnamefont {H.~J.}\ \bibnamefont
  {Warringa}},\ }\href {\doibase 10.1016/j.nuclphysa.2010.02.003} {\bibfield
  {journal} {\bibinfo  {journal} {Nucl. Phys.}\ }\textbf {\bibinfo {volume}
  {A836}},\ \bibinfo {pages} {311} (\bibinfo {year} {2010})},\ \Eprint
  {http://arxiv.org/abs/0912.2961} {arXiv:0912.2961 [hep-ph]} \BibitemShut
  {NoStop}%
\bibitem [{\citenamefont {Fukushima}(2013)}]{Fukushima:2012vr}%
  \BibitemOpen
  \bibfield  {author} {\bibinfo {author} {\bibfnamefont {K.}~\bibnamefont
  {Fukushima}},\ }\href {\doibase 10.1007/978-3-642-37305-3_9} {\bibfield
  {journal} {\bibinfo  {journal} {Lect. Notes Phys.}\ }\textbf {\bibinfo
  {volume} {871}},\ \bibinfo {pages} {241} (\bibinfo {year} {2013})},\ \Eprint
  {http://arxiv.org/abs/1209.5064} {arXiv:1209.5064 [hep-ph]} \BibitemShut
  {NoStop}%
\end{thebibliography}%

\end{document}